\documentclass[11pt]{article}

\usepackage[letterpaper,margin=1.25in]{geometry}
\usepackage[T1]{fontenc}
\usepackage{microtype}
\usepackage[style=authoryear,giveninits=true,uniquename=false,uniquelist=false,maxbibnames=99]{biblatex}
\usepackage[title]{appendix}
\usepackage[colorlinks=true,citecolor=blue,urlcolor=blue]{hyperref}
\usepackage{amsmath}
\usepackage{amsthm}
\usepackage{amsfonts}
\usepackage{tikz}
\usepackage[ruled]{algorithm2e}
\usepackage[labelfont=bf]{caption}
\usepackage{booktabs}
\usepackage{threeparttable}
\usepackage{authblk}
\usepackage{nicematrix}
\usepackage[style=british]{csquotes}

\renewbibmacro{in:}{}
\DeclareNameAlias{sortname}{family-given}
\DeclareNameAlias{default}{family-given}

\newcommand*{\email}[1]{\href{mailto:#1}{\nolinkurl{#1}}}
\DeclareMathOperator{\argmin}{arg\,min}
\DeclareMathOperator{\e}{E}
\DeclareMathOperator{\var}{Var}
\DeclareMathOperator{\cov}{Cov}
\DeclareMathOperator{\p}{Pr}
\DeclareMathOperator{\dir}{Dirichlet}
\DeclareMathOperator{\dirp}{DP}
\DeclareMathOperator{\sign}{sign}
\newtheorem{lemma}{Lemma}
\newtheorem{proposition}{Proposition}
\newtheorem{theorem}{Theorem}

\numberwithin{equation}{section}

\newcommand\keywords[1]{%
\begin{NoHyper}
\renewcommand\thefootnote{}\footnote{\emph{Keywords:} #1}%
\addtocounter{footnote}{-1}%
\end{NoHyper}
}

\title{Familial inference: tests for hypotheses on a family of centres}
\author[1]{Ryan Thompson\thanks{Corresponding author. Now at School of Mathematics and Statistics, University of New South Wales and Data61, Commonwealth Scientific and Industrial Research Organisation. Email: \email{ryan.thompson1@unsw.edu.au}}}
\author[1]{Catherine S. Forbes}
\author[2]{Steven N. MacEachern}
\author[2]{Mario Peruggia}
\affil[1]{Department of Econometrics and Business Statistics, Monash University}
\affil[2]{Department of Statistics, The Ohio State University}

\begin{document}

\maketitle

\begin{abstract}
Statistical hypotheses are translations of scientific hypotheses into statements about one or more distributions, often concerning their centre. Tests that assess statistical hypotheses of centre implicitly assume a specific centre, e.g., the mean or median. Yet, scientific hypotheses do not always specify a particular centre. This ambiguity leaves the possibility for a gap between scientific theory and statistical practice that can lead to rejection of a true null. In the face of replicability crises in many scientific disciplines, significant results of this kind are concerning. Rather than testing a single centre, this paper proposes testing a family of plausible centres, such as that induced by the Huber loss function (the Huber family). Each centre in the family generates a testing problem, and the resulting family of hypotheses constitutes a familial hypothesis. A Bayesian nonparametric procedure is devised to test familial hypotheses, enabled by a novel pathwise optimization routine to fit the Huber family. The favourable properties of the new test are demonstrated theoretically and experimentally. Two examples from psychology serve as real-world case studies.
\end{abstract}


\section{Introduction}
\label{sec:intro}

Hypothesis testing is one of statistics' most important contributions to the scientific method. Testing helps advance diverse lines of inquiry, from evaluating the efficacy of experimental drugs to assessing the validity of psychological theories. Researchers working on these problems often characterize their questions as competing statements about a centre $\mu$ of one or more distributions. In the simplest one-sample setting, these statements take the form
\begin{equation*}
\mathrm{H}_0:\mu\in\mathcal{M}_0\quad\text{vs.}\quad\mathrm{H}_1:\mu\in\mathcal{M}_1,
\end{equation*}
where $\mathcal{M}_0$ and $\mathcal{M}_1$ are a partition of the support $\mathcal{M}$ of $\mu$. There are myriads of classical tests for one- and two-sample hypotheses of centre. When $\mu$ is the mean, the most well-known of these is the $t$ test \parencite{Student1908}, and its extension to independent samples from populations with differing variances \parencite{Welch1947}. When $\mu$ is the median, the sign test \parencite{Fisher1925} is available, as is the median test for independent samples \parencite{Mood1950}. The signed-rank test, or rank-sum test for independent samples, are also tests of medians under certain assumptions \parencite{Wilcoxon1945,Mann1947}.

The possibility to test different centres such as the mean and median raises the question of what qualifies as a centre. We posit that a centre of a random variable $X$ should satisfy at least two criteria: (1) a reflection of $X$ about the centre should preserve the centre, and (2) a shift in $X$ by a constant should move the centre by that same constant. This definition is purposefully broad to accommodate the many notions of centre used throughout statistics. The mean and median trivially satisfy these criteria, as do other popular notions such as the mode, trimmed mean, and Winsorized mean. Quantiles other than the median, and by extension order statistics such as the minimum and maximum, are not centres under these criteria as they are not preserved by reflection in general. Still, the fact that there are many possibilities for centre can complicate hypothesis testing in science.

In certain applied areas, e.g., psychology and medicine, \emph{scientific hypotheses} are often silent about a specific centre and instead tend to be statistically vague, e.g., treatment A is more efficacious than treatment B. This ambiguity makes translation to \emph{statistical hypotheses} inherently subjective and can leave researchers questioning which centre to use. See \textcite{Blakely2001}, \textcite{BenAharon2019}, and \textcite{Rousselet2020} for discussions of this issue in epidemiology, medicine, and psychology. Moreover, ambiguity about the correct or best centre leaves the possibility for a gap between scientific theory and statistical practice that can lead to rejection of a true null, threatening the validity of findings. Sometimes $\mathrm{H}_0$ can be rejected just by switching from one centre to another, say from the mean to the median. In the face of replicability crises in various disciplines \parencite[see, e.g.,][]{Ioannidis2005,Open2015,Christensen2018}, the possibility for significant results of this sort is concerning. The fact that statistical experts often have no input on the statistical aspects of scientific research only aggravates the issue \parencite[see, e.g.,][]{Strasak2007,Hardwicke2020}. Transparent statistical tools are needed to instil confidence in scientific claims.

Motivated by the preceding discussion, this paper proposes a new approach to hypothesis testing: familial inference. Unlike existing inferential methods, which test hypotheses about a single centre, methods for familial inference test hypotheses about a \emph{family} of plausible centres, with the ultimate goal of strengthening any claims of significance. More specifically, consider a family of centres $\{\mu(\lambda):\lambda\in\Lambda\}$ where $\lambda$ indexes each member (centre). The familial testing problem is to decide which hypothesis concerning this family is correct:
\begin{equation*}
\mathrm{H}_0:\mu(\lambda)\in\mathcal{M}_0\text{ for some }\lambda\in\Lambda\quad\text{vs.}\quad\mathrm{H}_1:\mu(\lambda)\in\mathcal{M}_1\text{ for all }\lambda\in\Lambda.
\end{equation*}
The familial null hypothesis states that at least one member (centre) of the family is contained in the null set $\mathcal{M}_0$. The alternative hypothesis is that no member is in $\mathcal{M}_0$.\footnote{This style of testing may be considered to have an intersection-union test format. See the discussion in Appendix~\ref{app:iutesting}.} This paper studies the family of centres induced by the Huber loss function \parencite{Huber1964}. The Huber function is a mixture of square and absolute loss, where $\lambda$ controls the mixture. By sweeping $\lambda$ between 0 and infinity, one obtains a family of centres that includes the mean and median as limit points. All members of this Huber family satisfy our criteria for centre. While this more conservative testing approach could potentially overlook a useful treatment, it reduces the risk of promoting an ineffective (or harmful) treatment, wasting limited resources, and misdirecting future research. Given the well-reported scientific reproducibility crises, the benefits may far outweigh the cost.

Familial inference is more sophisticated than inference for a single centre and requires new tools developed in this paper. Our first methodological development is a Bayesian nonparametric procedure for one- and two-sample testing with continuous and discrete random variables. The procedure is based on the limit of a Dirichlet process prior \parencite{Ferguson1973}, sometimes referred to as the Bayesian bootstrap \parencite{Rubin1981}. Bayesian tests have several advantages over frequentist tests, including that they measure the probability of $\mathrm{H}_0$. Unlike $p$-values, Bayesian probabilities directly quantify uncertainty, arguably providing a clearer perspective on evidence for practitioners. We refer the reader to \textcite{Kruschke2013} and \textcite{Benavoli2017} for discussions on the merits of Bayesian testing. Besides the advantages of a Bayesian approach, the nonparametric nature of our test ameliorates concern about model misspecification. Though numerous existing works address Bayesian nonparametric testing \parencite{Ma2011,Benavoli2014,Huang2014,Benavoli2015,Holmes2015,Filippi2017,Gutierrez2019,Pereira2020}, these treat hypotheses about single statistical parameters or entire distributions, distinct from the familial hypotheses treated in this paper.

Our second methodological development is an algorithm for fitting the Huber family, necessary to implement the new test. The algorithm is a pathwise optimization routine that exploits piecewise linearity of the Huber solution path to fit the family, containing infinitely many centres, in a single pass over the data. It has low computational complexity and terminates in at most $n-1$ steps, where $n$ is the sample size. We elucidate the connection between our algorithm and least angle regression \parencite{Efron2004,Rosset2007}, popularly used for fitting the lasso regularization path \parencite{Tibshirani1996}. The algorithms devised in this paper are made available in the open-source \texttt{R} package \texttt{familial}, designed with a standard interface similar to that of existing tests in the \texttt{stats} package. Methods for visualizing the posterior family via functional boxplots \parencite{Sun2011} are provided. \texttt{familial} is publicly available on the \texttt{R} repository \texttt{CRAN}.

\begin{figure}[ht!]
\centering
\includegraphics{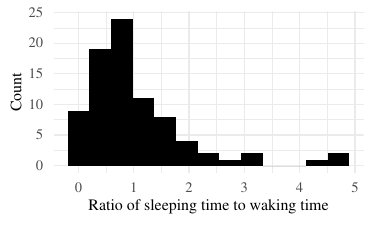}
\caption{Histogram of the mammalian sleep data.}
\label{fig:mammalian-sleep-histogram}
\end{figure}

To illustrate our proposal, we consider data from a study of mammalian sleep patterns in \textcite{Savage2007}. The data contains sleep times for $n=83$ species of mammals. A histogram of the ratio of sleeping hours to waking hours is plotted in Figure~\ref{fig:mammalian-sleep-histogram}. The data are heavily right-skewed, suggesting the mean and median are probably far separated.
Suppose we ask whether mammals tend to spend as much time sleeping as they do awake, i.e., whether $\mu=1$. A $t$ test that the mean is one yields a $p$-value of 0.698. A bootstrap test that the mean is one, conducted as a robustness check, produces only a marginally smaller $p$-value of 0.668. A sign test that the median is one gives a $p$-value of 0.028. At a conventional 0.05 significance level, these tests do not yield the same answer to our scientific question. This inconsistency raises the question of how exactly to proceed in the absence of a guiding scientific theory.

\begin{figure}[ht!]
\centering
\includegraphics{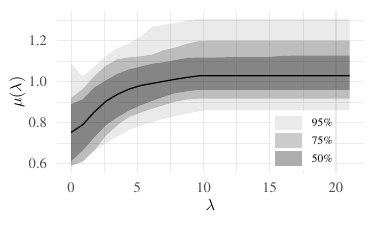}
\caption{Functional boxplot of the posterior Huber family for the mammalian sleep data. Shading indicates different central regions of the posterior.}
\label{fig:mammalian-sleep-boxplot}
\end{figure}

Using our procedure, we estimate the posterior Huber family via 1,000 Bayesian bootstraps, summarized in Figure~\ref{fig:mammalian-sleep-boxplot} by a functional boxplot.
As the Huber parameter $\lambda\to\infty$, the 50\% central region of the posterior encloses the null value (recall the mean is attained in the limit). By querying the posterior, we find a probability of 0.633 that at least one centre in the family equals one. Under zero-one loss configured analogously to using a 0.05 frequentist significance level (detailed later), the familial test finds insufficient evidence to reject the null in favour of the alternative. Because no specific choice was made about the centre, the problem of choosing between conflicting tests does not arise. Most importantly, we do not arrive at a result that would hold only under a certain centre.

Before delving into the details of our approach, we pause to ask whether a familial test of centres might be less useful than a more general test of distributions. Earlier Bayesian nonparametric tests of distributions, such as those of \textcite{Holmes2015} and \textcite{Pereira2020}, consider a larger class of alternatives than ours. The usefulness of these tests depends, however, upon the scientific question. A change in scale, e.g., can cause them to reject even with equal centres. If such changes are not scientifically pertinent, these tests are unsuitable. In contrast, our familial test directly identifies discrepancies in centres, something which alternative tests overlook or conflate with other distributional attributes.

\section{Bayesian nonparametric test}
\label{sec:test}

\subsection{Inference problem}

Let $X_1,\ldots,X_n$ be an iid sample according to a distribution $P_0$. Our goal is to carry out inference on the set $\{\mu_0(\lambda):\lambda\in\Lambda\}$, where
\begin{equation*}
\mu_0(\lambda):=\underset{\mu\in\mathcal{M}}{\argmin}\e\left[\ell_\lambda\left(\frac{X-\mu}{\sigma}\right)\right]=\underset{\mu\in\mathcal{M}}{\argmin}\int\ell_\lambda\left(\frac{x-\mu}{\sigma}\right)\,dP_0(x).
\end{equation*}
Here, $\ell_\lambda:\mathbb{R}\to\mathbb{R}_+$ is a loss function controlled by the parameter $\lambda$. The constant $\sigma>0$ is necessary in certain loss functions to make $\lambda$ invariant to the spread of $X$. Invariance of $\lambda$ to spread is relevant for testing independent samples, addressed later. The population centre $\mu_0(\lambda)$ minimizes the expectation of the loss configured by $\lambda$ under $P_0$. To maintain generality throughout this section, we do not specify a particular loss function. However, to give a concrete example that will be the focus of subsequent sections, one may consider the Huber function
\begin{equation*}
\ell_\lambda(z)=
\begin{cases}
\frac{1}{2}z^2, & \text{if }|z|<\lambda, \\
\lambda|z|-\frac{1}{2}\lambda^2, & \text{if }|z|\geq\lambda.
\end{cases}
\end{equation*}
The support of $\lambda$ is $\Lambda=(0,\infty)$. The mean of $P_0$ is the limiting solution as $\lambda\to\infty$. The median is the limiting solution in the other direction. The continuum of centres therebetween comprises the Huber family. The approach we propose can accommodate the restriction to any subset of the family given by $\Lambda=[a,b]$ for $0<a<b<\infty$.

If the true generative model $P_0$ were known, we would immediately have access to the family $\{\mu_0(\lambda):\lambda\in\Lambda\}$. Of course, this is not the case in practice, $P_0$ is unknown. The traditional parametric Bayesian approach to this problem proceeds by means of a prior on parameters for a class of models for $P$. A valid criticism of this approach is the implicit assumption that $P_0$ is contained in the model class. Misspecified models can lead to false conclusions, which is troubling in the context of hypothesis testing. To this end, the Bayesian nonparametric approach is an appealing alternative. Rather than placing a prior on the parameters governing a distribution for $P$, one places a prior directly on the distribution itself. The Dirichlet process, a probability distribution on the space of probability distributions, is a natural candidate for this task. Since Dirichlet processes have support on a large class of distributions, they are a popular prior in Bayesian nonparametrics. The reader is referred to \textcite{MacEachern2016} for a recent and accessible overview of their properties.

\subsection{Bayesian bootstrap}

We denote by $\dirp(cP_\pi)$ a Dirichlet process with base distribution $P_\pi$ and concentration parameter $c>0$. The concentration parameter is used to impart confidence in $P_\pi$. With a Dirichlet process as a prior on $P$, our Bayesian model is
\begin{equation*}
X_1,\ldots,X_n\,|\,P\overset{\mathrm{iid}}{\sim}P,\quad P\sim\dirp(cP_\pi).
\end{equation*}
\textcite{Ferguson1973} shows the posterior corresponding to this model is also a Dirichlet process:
\begin{equation*}
P\,|\,X_1=x_1,\ldots,X_n=x_n\sim\dirp\left(cP_\pi+\sum_{i=1}^n\delta_{x_i}\right),
\end{equation*}
where $\delta_{x_i}$ is the Dirac measure at $x_i$. The Dirichlet process is a conjugate prior for iid sampling under $P$, and the posterior is the base distribution $P_\pi$ with added point masses at the sample realizations $x_1,\ldots,x_n$. A base distribution $P_\pi$ must be chosen to operationalize this model. If one wishes to minimize the impact of the choice of $P_\pi$, it is sensible to consider the limiting case where the concentration parameter $c\to0$, which leads to the posterior
\begin{equation*}
P\,|\,X_1=x_1,\ldots,X_n=x_n\sim\dirp\left(\sum_{i=1}^n\delta_{x_i}\right).
\end{equation*}
\textcite{Gasparini1995} shows that this posterior exactly matches the Bayesian bootstrap, proposed by \textcite{Rubin1981} as the Bayesian analog of the frequentist bootstrap \parencite{Efron1979}. \textcite{MacEachern1993} also establishes a unique connection of this posterior to the empirical distribution of the data. The Bayesian bootstrap places support only on the observed data and is equivalent to
\begin{equation*}
P(\cdot)=\sum_{i=1}^nw_i\delta_{x_i}(\cdot),\quad(w_1,\ldots,w_n)\sim\dir(1,\ldots,1),
\end{equation*}
where $\dir(1,\ldots,1)$ is the $n$-dimensional Dirichlet distribution with all concentration parameters equal to one. Sometimes this distribution is referred to as flat or uniform. The first- and second-order asymptotic properties of the Bayesian bootstrap are described in \textcite{Lo1987} and \textcite{Weng1989}. As well as being theoretically well-understood, the Bayesian bootstrap admits scalable sampling algorithms that are trivially parallelizable, making posterior exploration highly tractable. See \textcite{Fong2019}, \textcite{Lyddon2019}, and \textcite{Barrientos2020} for recent applications of the Bayesian bootstrap to complex models and data. As with those applications, tractability is key here.

We now have a posterior for $P$, and consequently also a posterior on any summaries of $P$ \parencite[see, e.g.,][]{Lee2014}, including those of interest: families of centres. To estimate the posterior for a given family we propose Algorithm~\ref{alg:bootstrap}.
\begin{algorithm}
\caption{Bayesian bootstrap for familial inference}
\label{alg:bootstrap}
\begin{tabbing}
\quad Input $(x_1,\ldots,x_n)$ \\
\quad For $b=1,\ldots,B$: \\
\quad\quad 1. Sample $(w_1^{(b)},\ldots,w_n^{(b)})$ from $\dir(1,\ldots,1)$ \\
\quad\quad 2. Compute $\mu^{(b)}(\lambda)=\argmin_{\mu\in\mathcal{M}}\sum_{i=1}^nw_i^{(b)}\ell_\lambda([x_i-\mu]/\sigma^{(b)})$ for all $\lambda\in\Lambda$ \\
\quad Output $\{\mu^{(b)}(\lambda):\lambda\in\Lambda\}_{b=1}^B$
\end{tabbing}
\vspace{-1em}
\end{algorithm}
The complexity of solving the minimization problem in step two for all $\lambda\in\Lambda$ depends on the loss function. In the next section, we present a numerical routine that addresses the case where the loss function is the Huber function. Since $\lambda$ in the Huber function is sensitive to changes in spread, we configure $\sigma^{(b)}$ to be the median absolute deviation of the bootstrap sample, i.e., the weighted median absolute deviation with weights $w_1^{(b)},\ldots,w_n^{(b)}$. The standard deviation of the bootstrap sample could also be used. In our experience, switching between standard deviation and median absolute deviation usually does not lead to materially different outcomes, and the choice is only relevant for independent samples testing (see Section~\ref{sec:two-sample}).

From the output of Algorithm~\ref{alg:bootstrap}, the posterior probabilities $p_{\mathrm{H}_0}:=\p(\mathrm{H}_0\,|\,x_1,\ldots,x_n)$ and $p_{\mathrm{H}_1}:=\p(\mathrm{H}_1\,|\,x_1,\ldots,x_n)$ are estimable as
\begin{equation*}
\hat{p}_{\mathrm{H}_0}:=\frac{1}{B}\sum_{b=1}^B1(\exists\,\lambda\in\Lambda:\mu^{(b)}(\lambda)\in\mathcal{M}_0)
\end{equation*}
and
\begin{equation*}
\hat{p}_{\mathrm{H}_1}:=\frac{1}{B}\sum_{b=1}^B1(\forall\,\lambda\in\Lambda:\mu^{(b)}(\lambda)\in\mathcal{M}_1).
\end{equation*}
Since $\mathrm{H}_0$ and $\mathrm{H}_1$ are mutually exclusive and collectively exhaustive, $p_{\mathrm{H}_0}+p_{\mathrm{H}_1}=1$ and, for any $B$, $\hat{p}_{\mathrm{H}_0}+\hat{p}_{\mathrm{H}_1}=1$.

Though our focus is the Bayesian bootstrap (i.e., the Dirichlet process prior with concentration parameter $c\to0$), one can extend Algorithm 1 to handle the prior with $c>0$ via the stick-breaking construction of \textcite{Sethuraman1994}. The interested reader may refer to Chapter 2 of \textcite{Muller2015} for details of this approach.

\subsection{Decision rule}

To map the estimated posterior probabilities $\hat{p}_{\mathrm{H}_0}$ and $\hat{p}_{\mathrm{H}_1}$ to a decision, we assign a loss to each possible decision. Specifically, given the posterior probability vector $\hat{p}=(\hat{p}_{\mathrm{H}_0},\hat{p}_{\mathrm{H}_1})^\top$, we make the decision giving lowest posterior expected loss $L\hat{p}$, where $L$ is loss matrix with rows corresponding to the decision to accept $\mathrm{H}_0$, accept $\mathrm{H}_1$, or accept neither (an \emph{indeterminate} decision). We use
\begin{equation}
\label{eq:lossmat}
L:=
\begin{pmatrix}
l_{\mathrm{H}_0|\mathrm{H}_0} & l_{\mathrm{H}_0|\mathrm{H}_1} \\
l_{\mathrm{H}_1|\mathrm{H}_0} & l_{\mathrm{H}_1|\mathrm{H}_1} \\
l_{\mathrm{I}|\mathrm{H}_0} & l_{\mathrm{I}|\mathrm{H}_1}
\end{pmatrix}
=
\begin{pNiceMatrix}[first-col,first-row,code-for-first-col=\scriptstyle,code-for-first-row=\scriptstyle]
& \mathrm{H}_0 & \mathrm{H}_1 \\
\mathrm{H}_0 & 0 & 20 \\
\mathrm{H}_1 & 20 & 0 \\
\mathrm{I} & 1 & 1 \\
\end{pNiceMatrix},
\end{equation}
where $l_{\mathrm{H}_j|\mathrm{H}_k}$ denotes the loss incurred in accepting $\mathrm{H}_j$ when $\mathrm{H}_k$ is true for $j,k=0,1$, and where $l_{\mathrm{I}|\mathrm{H}_k}$ denotes the loss from an indeterminate decision for $k=0,1$. Under the above configuration of $L$, either $\mathrm{H}_0$ or $\mathrm{H}_1$ is accepted depending on whether $\hat{p}_{\mathrm{H}_0}$ or $\hat{p}_{\mathrm{H}_1}$ is greater than 0.95, analogous to a 0.05 level frequentist test. When both probabilities are less than 0.95 the decision is indeterminate.

\subsection{Two-sample problem}
\label{sec:two-sample}

The discussion up to now has focused on the one-sample setting. Consider now the two-sample setting with samples $X_1,\ldots,X_{n_1}$ and $Y_1,\ldots,Y_{n_2}$. If $X_i$ and $Y_i$ are meaningfully coupled together, e.g., measurements on the same subject before and after treatment, the two samples are paired and $n_1=n_2$. Define the random variable $Z_i$ as the difference $X_i-Y_i$. Then the familial hypotheses are
\begin{equation*}
\mathrm{H}_0:\mu_Z(\lambda)\in\mathcal{M}_0\text{ for some }\lambda\in\Lambda\quad\text{vs.}\quad\mathrm{H}_1:\mu_Z(\lambda)\in\mathcal{M}_1\text{ for all }\lambda\in\Lambda,
\end{equation*}
where $\mu_Z(\lambda)$ is a centre of $Z$. Algorithm \ref{alg:bootstrap} applies directly to the sample $Z_1,\ldots,Z_n$ with $n=n_1=n_2$.

When $X_i$ and $Y_i$ are not coupled together the samples are independent. The familial hypotheses are then
\begin{equation*}
\let\scriptstyle\textstyle
\mathrm{H}_0:\substack{\mu_X(\lambda)-\mu_Y(\lambda)\in\mathcal{M}_0 \\ \text{ for some }\lambda\in\Lambda}\quad\text{vs.}\quad\mathrm{H}_1:\substack{\mu_X(\lambda)-\mu_Y(\lambda)\in\mathcal{M}_1 \\ \text{ for all }\lambda\in\Lambda.}
\end{equation*}
Here, the same centre of $X$ is compared with the same centre of $Y$, i.e., the mean is compared with the mean, the median with the median, and so on. Testing these hypotheses requires bootstrapping the families of $X$ and $Y$ with independently drawn weights. When independent weights are drawn at a bootstrap iteration, each centre of $Y$ is subtracted from the same centre of $X$. The posterior probability of $\mathrm{H}_0$ is estimated by the proportion of times across bootstrap iterations that the set of differences intersects the null set $\mathcal{M}_0$.

\section{Huber family}
\label{sec:huber}

\subsection{Optimization problem}

To implement the testing procedure of the preceding section, we require a method for fitting the family of centres to each distribution drawn from the posterior, i.e., for solving the optimization problems in step two of Algorithm~\ref{alg:bootstrap} given fixed bootstrap weights $w_1^{(b)},\ldots,w_n^{(b)}$. For simplicity of exposition, we drop the bootstrap iteration superscript $(b)$ and fix $\sigma=1$ without loss of generality. The Huber function as a function of the residual $x-\mu$ can then be expressed as
\begin{equation*}
\ell_\lambda(x-\mu)=
\begin{cases}
\frac{1}{2}(x-\mu)^2, & \text{if }|x-\mu|<\lambda, \\
\lambda|x-\mu|-\frac{1}{2}\lambda^2, & \text{if }|x-\mu|\geq\lambda.
\end{cases}
\end{equation*}
We denote the loss over the weighted (bootstrap) sample by
\begin{equation*}
\mathcal{L}_\lambda(\mu):=\sum_{i=1}^nw_i\ell_\lambda(x_i-\mu).
\end{equation*}
Our goal is to devise an algorithm for computing the set $\{\mu(\lambda):\lambda\in\Lambda\}$, where
\begin{equation}
\label{eq:huberopt}
\mu(\lambda):=\underset{\mu\in\mathbb{R}}{\argmin}\,\mathcal{L}_\lambda(\mu)
\end{equation}
and $\Lambda=(0,\infty)$. For an equally weighted sample, \eqref{eq:huberopt} includes as limiting cases the sample mean and sample median. When the weights are unequal, the limit points become the \emph{weighted mean} and \emph{weighted median}, interpretable as the mean and median of the bootstrap sample. The weighted mean is defined by
\begin{equation*}
\bar{\mu}:=\underset{\mu\in\mathbb{R}}{\argmin}\,\sum_{i=1}^nw_i(x_i-\mu)^2=\sum_{i=1}^nw_ix_i,
\end{equation*}
and the weighted median by
\begin{equation*}
\tilde{\mu}:=\underset{\mu\in\mathbb{R}}{\argmin}\,\sum_{i=1}^nw_i|x_i-\mu|.
\end{equation*}
There is no analytical solution for the weighted median. In fact, the weighted mean is the only Huber centre that admits an analytical solution in general. For general $\lambda$, the existence of a solution to the minimization \eqref{alg:pathwise} requires only that the realized sample $x_1,\ldots,x_n$ and weights $w_1,\ldots,w_n$ be finite. Hence, successful computation of a solution does not hinge upon the existence of a solution in the population, e.g., if $x_1,\ldots,x_n$ are realizations of a Cauchy.

If $\Lambda$ were a finite set, it would be possible to solve the optimization problem~\eqref{eq:huberopt} for each of its elements. For given $\lambda$, the one-dimensional problem~\eqref{eq:huberopt} is convex, and although it does not admit an analytical solution, it is amenable to simple numerical routines \parencite{Huber2009}. Even if $\Lambda$ is not finite, one might try approximating it using a fine grid and then proceed to solve each minimization individually. Recall though each set of minimization problems needs to be solved $B$ times in the Bayesian bootstrap, where $B$ might be 1,000, 10,000, or larger. Thus, even with an efficient algorithm, total cumulative runtime can be prohibitive. Notwithstanding runtime considerations, such an approach still only yields an approximation. Instead of an approximation, we propose a fast and exact pathwise algorithm that optimizes \eqref{eq:huberopt} for all values of $\lambda$.

\subsection{Pathwise optimization routine}

Our approach exploits piecewise linearity of the solution path $\mu(\lambda)$ for $\lambda\in(0,\infty)$, a property we now demonstrate. The gradient of the Huber function with respect to $\mu$ is
\begin{equation*}
\begin{split}
\frac{\partial\ell_\lambda(x-\mu)}{\partial\mu}&=
\begin{cases}
-(x-\mu), & \text{if } |x-\mu|<\lambda, \\
-\lambda\sign(x_i-\mu), & \text{if } |x-\mu|\geq\lambda.
\end{cases}
\end{split}
\end{equation*}
Hence, the gradient of the loss over the weighted sample is
\begin{equation*}
\frac{\partial\mathcal{L}_\lambda(\mu)}{\partial\mu}=\sum_{i=1}^nw_i\frac{\partial\ell_\lambda(x_i-\mu)}{\partial\mu}=-\sum_{i:|x_i-\mu|<\lambda}w_i(x_i-\mu)-\sum_{i:|x_i-\mu|\geq\lambda}w_i\lambda\sign(x_i-\mu).
\end{equation*}
We denote the above gradient by $\mathcal{L}'(\mu)$, suppressing the explicit dependency on $\lambda$. The first-order condition for optimality of $\mu(\lambda)$ requires that $\mathcal{L}'(\mu(\lambda))=0$. The implicit function theorem then gives
\begin{equation*}
\frac{\partial\mu(\lambda)}{\partial\lambda}=-\frac{\partial\mathcal{L}'(\mu(\lambda))}{\partial\lambda}/\left.\frac{\partial\mathcal{L}'(\mu)}{\partial\mu}\right|_{\mu=\mu(\lambda)},
\end{equation*}\
which, after evaluating gradients on the right-hand side, yields
\begin{equation}
\label{eq:gradient}
\frac{\partial\mu(\lambda)}{\partial\lambda}=\frac{\sum_{i:|x_i-\mu(\lambda)|\geq\lambda}w_i\sign(x_i-\mu(\lambda))}{\sum_{i:|x_i-\mu(\lambda)|<\lambda}w_i}.
\end{equation}
Observe that the gradient of the solution path $\partial\mu(\lambda)/\partial\lambda$ is piecewise constant as a function of $\lambda$, implying that $\mu(\lambda)$ is piecewise linear. It follows that $\mu(\lambda)$ is also piecewise continuous with left and right limits. It can be verified that the left and right limits at any knot $\lambda^\star$ equal $\mu(\lambda^\star)$, and hence that $\mu(\lambda)$ is continuous.

Since the solution path is piecewise linear, it is composed of a sequence of knots, i.e., certain values of $\lambda$ at which $|x_i-\mu(\lambda)|=\lambda$ for one or more sample points. These knots correspond to crossing events, where sample points transition between the square and absolute pieces of the Huber function. Lemma~\ref{lemma:crossing} characterizes a useful property in relation to these crossing events.
\begin{lemma}
\label{lemma:crossing}
Suppose sample point $x_0$ satisfies $|x_0-\mu(\lambda^\star)|\geq\lambda^\star$ for some $\lambda^\star>0$. Then, for all $0<\lambda<\lambda^\star$, it holds $|x_0-\mu(\lambda)|\geq\lambda$.
\end{lemma}
Lemma~\ref{lemma:crossing} implies that, for a decreasing sequence of $\lambda$, once a sample point has crossed to the absolute piece of the Huber function, it remains there. This property guarantees the existence of at most $n$ knots along the solution path. The lemma is proven in Appendix~\ref{app:huber}.

To trace out the solution path, we need only fit $\mu$ at each $\lambda$ in the sequence of knots since any solution between knots is linearly interpolable. A method to efficiently determine the location and solution at each knot is required. Suppose we are at an arbitrary point $(\lambda,\mu)$ along the solution path. Then, thanks to piecewise linearity, the closest knot point $(\lambda^+,\mu^+)$ to the left of $(\lambda,\mu)$ is computable by taking a step $\gamma>0$, of a certain size, as follows:
\begin{equation}
\label{eq:lambda+}
\lambda^+=\lambda-\gamma
\end{equation}
and
\begin{equation}
\label{eq:mu+}
\mu^+=\mu-\gamma\frac{\partial\mu(\lambda)}{\partial\lambda}.
\end{equation}
Equation~\eqref{eq:gradient} provides an analytical expression for the gradient $\partial\mu(\lambda)/\partial\lambda$. An expression for the required step size $\gamma$ is still needed. To this end, we present Proposition~\ref{prop:step}.
\begin{proposition}
\label{prop:step}
Let $(\lambda,\mu)$ be any point along the solution path such that $|x_i-\mu|<\lambda$ for at least one $i=1,\ldots,n$. Then the largest positive step size before the solution path reaches a knot point $(\lambda^+,\mu^+)$ to the left of $(\lambda,\mu)$ is
\begin{equation*}
\gamma=\min_{i:|x_i-\mu|<\lambda}\left(\frac{\lambda-s_i(x_i-\mu)}{1+s_i\partial\mu(\lambda)/\partial\lambda}\right),
\end{equation*}
where $s_i=\sign(x_i-\tilde{\mu})$ and $\tilde{\mu}$ is the weighted median.
\end{proposition}
The requirement $|x_i-\mu|<\lambda$ for at least one $i=1,\ldots,n$ guarantees the existence of at least one more unexplored knot along the solution path. Beyond the first and last knots, the solution path is flat. The proposition is proven in Appendix~\ref{app:huber}.

\begin{figure}[ht!]
\centering
\includegraphics{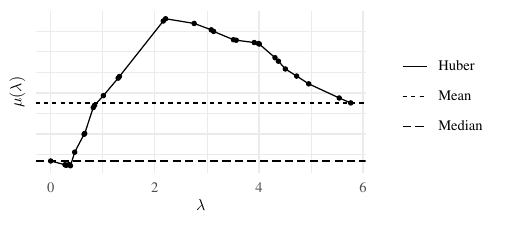}
\caption{Algorithm~\ref{alg:pathwise} applied with $x_1,\ldots,x_n$ drawn from a standard normal distribution and $w_1,\ldots,w_n$ drawn from a flat Dirichlet distribution with $n=30$. The solid points are iterates (knots) from the algorithm. Centres between iterates are linearly interpolated.}
\label{fig:pathwise}
\end{figure}

Putting together the above ingredients and letting $\lambda=\lambda^{(m)}$, $\lambda^+=\lambda^{(m+1)}$, $\mu=\mu^{(m)}$, and $\mu^+=\mu^{(m+1)}$ we arrive at Algorithm~\ref{alg:pathwise}.
\begin{algorithm}
\caption{Pathwise optimization for the Huber family}
\label{alg:pathwise}
\begin{tabbing}
\quad Input $(x_1,\ldots,x_n)$ and $(w_1,\ldots,w_n)$ \\
\quad Initialize $\mu^{(1)}=\sum_{i=1}^nw_ix_i$ and $\lambda^{(1)}=\max_{i}(|x_i-\mu^{(1)}|)$ \\
\quad 1. Calculate the sign $s_i=\sign(x_i-\tilde{\mu})$ for $i=1,\ldots,n$ \\
\quad For $m=1,\ldots,n-1$: \\
\quad\quad 2. If $\{i:|x_i-\mu^{(m)}|<\lambda^{(m)}\}=\emptyset$ then $m=m-1$ and break \\
\quad\quad 3. Calculate the gradient
\end{tabbing}
\begin{equation*}
\eta=\frac{\sum_{i:|x_i-\mu^{(m)}|\geq\lambda^{(m)}}w_i\sign(x_i-\mu^{(m)})}{\sum_{i:|x_i-\mu^{(m)}|<\lambda^{(m)}}w_i}
\end{equation*}
\begin{tabbing}
\quad\qquad 4. Calculate the step size
\end{tabbing}
\begin{equation*}
\gamma=\min_{i:|x_i-\mu^{(m)}|<\lambda^{(m)}}\left(\frac{\lambda^{(m)}-s_i(x_i-\mu^{(m)})}{1+s_i\eta}\right)
\end{equation*}
\begin{tabbing}
\quad\quad 5. Perform the updates $\lambda^{(m+1)}=\lambda^{(m)}-\gamma$ and $\mu^{(m+1)}=\mu^{(m)}-\gamma\eta$ \\
\quad Output $(\lambda^{(1)},\ldots,\lambda^{(m+1)})$ and $(\mu^{(1)},\ldots,\mu^{(m+1)})$
\end{tabbing}
\vspace{-1em}
\end{algorithm}
Starting at the rightmost knot point $(\lambda^{(1)},\mu^{(1)})$, which corresponds to the weighted mean, the algorithm forges a path step-by-step to the leftmost knot point, which corresponds to the weighted median. Figure~\ref{fig:pathwise} illustrates this process on $n=30$ iid draws from a standard normal distribution.
The algorithm begins at a value of $\lambda$ large enough to induce the weighted mean as the centre and then iteratively decreases $\lambda$. The final $\lambda$ in this sequence of iterates is sufficiently small to induce the weighted median as the centre. Observe that the path is piecewise linear and continuous.

Thus far the spread has been fixed at $\sigma=1$. To recover the solution path for $\sigma\neq1$, we scale the output $(\lambda^{(1)},\ldots,\lambda^{(m+1)})$ from Algorithm~\ref{alg:pathwise} by multiplying it by $\sigma$. The centres $(\mu^{(1)},\ldots,\mu^{(m+1)})$ do not change. This scaling has the intended effect of using the scaled residual $(x-\mu)/\sigma$ in the Huber function instead of $x-\mu$. We remind the reader that this scaling makes the solution path scale-free, relevant for testing independent samples.

\subsection{Relation to least angle regression}

Algorithm~\ref{alg:pathwise} bears similarity to least angle regression \parencite{Efron2004,Rosset2007}, a pathwise optimization routine that traces the solution path of lasso regression coefficients. To clarify this similarity, first recall the Moreau envelope $f_\lambda(z)$ of a real-valued function $f(z)$, which is the infimal convolution of $f(z)$ and $g(z)=1/(2\lambda)z^2$ \parencite[see, e.g.,][]{Polson2015}. When $f(z)=|z|$, there is a precise relation between the Huber function $\ell_\lambda(z)$ and the Moreau envelope:
\begin{equation*}
f_\lambda(z):=\underset{\beta\in\mathbb{R}}{\inf}\left(|\beta|+\frac{1}{2\lambda}(z-\beta)^2\right)=
\begin{cases}
\frac{1}{2\lambda}z^2, & \text{if } |z|<\lambda, \\
|z|-\frac{1}{2}\lambda, & \text{if } |z|\geq\lambda.
\end{cases}
\end{equation*}
The right-hand side is equal to $\ell_\lambda(z)/\lambda$. In words, multiplying the Moreau envelope of the absolute value function by $\lambda$ yields the Huber function, a known result from convex analysis \parencite{Beck2017}. Hence, we have the chain of equalities
\begin{equation*}
\begin{split}
\underset{\mu\in\mathbb{R}}{\min}\sum_{i=1}^nw_i\ell_\lambda(x_i-\mu)&=\underset{\mu\in\mathbb{R}}{\min}\,\sum_{i=1}^nw_i\lambda f_\lambda(x_i-\mu) \\
&=\underset{\mu\in\mathbb{R}}{\min}\sum_{i=1}^nw_i\underset{\beta_i\in\mathbb{R}}{\inf}\left(\frac{1}{2}(x_i-\mu-\beta_i)^2+\lambda|\beta_i|\right) \\
&=\underset{\mu,\beta_1,\ldots,\beta_n\in\mathbb{R}}{\min}\,\sum_{i=1}^nw_i\left(\frac{1}{2}(x_i-\mu-\beta_i)^2+\lambda|\beta_i|\right).
\end{split}
\end{equation*}
The infimum can be written as a minimum since the absolute value function is closed convex. The final line is a weighted lasso regression of $x_1,\ldots,x_n$ on an identity design matrix of dimensions $n\times n$, showing that the Huber problem~\eqref{eq:huberopt} can be recast as a weighted lasso problem. Thus, applying least angle regression, configured with weights, to an identity design matrix yields a path identical to that produced by Algorithm~\ref{alg:pathwise}. Despite this equivalence, the development of Algorithm~\ref{alg:pathwise} remains essential. Least angle regression is designed for general design matrices and, as such, does not exploit the structure of regression with an identity design, i.e., the Huber problem. Algorithm~\ref{alg:pathwise}, on the other hand, takes full advantage of this structure. In numerical experimentation, we observed that Algorithm~\ref{alg:pathwise} is typically an order of magnitude faster than least angle regression. Without this speedup, the Bayesian bootstrap would remain computationally burdensome.

\section{Consistency}

The asymptotic properties of the familial test are now established. We focus on a real-valued null $m_0$ in the one-sample (and paired samples) setting, which serves as a fundamental scenario, though our result is adaptable to a set-valued null $\mathcal{M}_0$ and independent samples. We consider two cases: (1) the null hypothesis $\mathrm{H}_0:\mu_0(\lambda)=m_0$ for some $\lambda\in\Lambda$ is true and (2) the alternative hypothesis $\mathrm{H}_1:\mu_0(\lambda)\neq m_0$ for all $\lambda\in\Lambda$ is true.

Theorem~\ref{thm:consistency} states the familial test is asymptotically consistent under both $\mathrm{H}_0$ and $\mathrm{H}_1$.
\begin{theorem}
\label{thm:consistency}
Let $X_1,\ldots,X_n$ be iid random variables with distribution $P_0$ and, independently, let $(w_1,\ldots,w_n)\sim\dir(1,\ldots,1)$. Let $\Lambda=[0,\bar{\lambda}]$ for some $0<\bar{\lambda}<\infty$. Define the functionals
\begin{equation*}
\hat{\mu}(\lambda):=\underset{\mu\in\mathbb{R}}{\argmin}\sum_{i=1}^nw_i\ell_\lambda(X_i-\mu)
\end{equation*}
and
\begin{equation*}
\mu_0(\lambda):=\underset{\mu\in\mathbb{R}}{\argmin}\int\ell_\lambda(x-\mu)\,dP_0(x).
\end{equation*}
Then the following results hold.
\begin{enumerate}
\item If $\mathrm{H}_0:\mu_0(\lambda)=m_0$ for some $\lambda\in\Lambda$ is true with $\mu_0(\lambda)$ strictly crossing through $m_0$, then
\begin{equation*}
\lim_{n\to\infty}\p(\exists\,\lambda\in\Lambda:\hat{\mu}(\lambda)=m_0)=1.
\end{equation*}
\item If $\mathrm{H}_1:\mu_0(\lambda)\neq m_0$ for all $\lambda\in\Lambda$ is true with $\mu_0(\lambda)$ bounded away from $m_0$, then
\begin{equation*}
\lim_{n\to\infty}\p(\forall\,\lambda\in\Lambda:\hat{\mu}(\lambda)\neq m_0)=1.
\end{equation*}
\end{enumerate}
\end{theorem}
A proof is available in Appendix~\ref{app:consistency}. The theorem requires only a single Bayesian bootstrap distribution, i.e., $B=1$, for consistency of the test. Having $B>1$ presents no technical complication, but it is unnecessary for consistency when $n\to\infty$.

At a high level, the proof of Theorem~\ref{thm:consistency} involves showing the estimator $\hat{\mu}(\lambda)$ (actually, its gradient) under a Bayesian bootstrap distribution can be made arbitrarily close to that under the true distribution $P_0$, uniformly over all $\lambda\in\Lambda$. This result allows us to argue if $\mathrm{H}_0$ is true and $\mu_0(\lambda)$ strictly crosses through $m_0$, then so does $\hat{\mu}(\lambda)$ in the limit. Alternatively, if $\mathrm{H}_0$ is false and $\mu_0(\lambda)$ is bounded above or below away from $m_0$, then so is $\hat{\mu}(\lambda)$ in the limit.

The case where the null is true and $\mu_0(\lambda)$ does not strictly cross through $m_0$ is not covered by Theorem~\ref{thm:consistency}. This case would occur, say, when $P_0$ is a normal distribution with mean $m_0$. However, symmetric distributions like the normal are uncommon in practice and not the focus of our test.

\section{Experiments}
\label{sec:case}

\subsection{Familial package}

This section reports experiments on real and synthetic data. To enable these exercises, the test and algorithms described in the preceding sections are implemented in the \texttt{R} package \texttt{familial}. For a sample of size $n=200$, \texttt{familial} takes about half a second to perform 1,000 bootstraps for a single sample on one core of a modern processor. Parallelism is also supported. Run time scales linearly with the sample size, number of bootstraps, and, if parallelised, number of processor cores.

\subsection{Body posture study}

\textcite{Rosenbaum2017} conducted an experiment to ascertain the effect of body posture on selective attention (refer to Experiment 3 in that paper). The experiment employed the Stroop test, where subjects are asked to announce colours of a sequence of words and not the words themselves, e.g., announce blue when the word red is printed in blue. The difference in response times between congruent word-colour pairs and incongruent pairs is the Stroop effect. Experimental subjects took the test once while sitting and once while standing. The study found standing lowered the Stroop effect compared with sitting, indicating improved selective attention while standing.

\begin{figure}[ht!]
\centering
\includegraphics{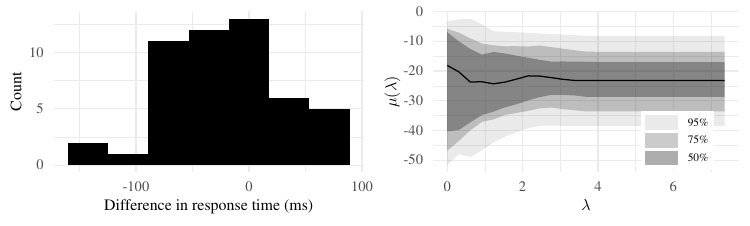}
\caption{Body posture data. The left plot is a histogram of the data. The right plot is a functional boxplot of the posterior Huber family. Shading indicates different central regions of the posterior.}
\label{fig:stroop}
\end{figure}

The dataset \parencite{Mama2018} contains paired observations on response times of $n=50$ subjects. Figure~\ref{fig:stroop} presents a histogram of differences in response time alongside a functional boxplot of the posterior Huber family.
The response times do not deviate markedly from a normal distribution, though they are slightly left-skewed. The posterior concentrates well below zero, suggesting standing might reduce the Stroop effect.

The study reported a $p$-value of 0.004 from an $F$ test of the interaction between congruency and posture in a repeated-measures ANOVA, equivalent to a Student $t$ test that the mean difference in response times is zero. The Fisher sign test and Wilcoxon signed-rank test produce $p$-values of 0.007 and 0.006, respectively. The Huber familial test finds that the probability of the null is 0.005. All tests reject the null that body posture does not affect the Stroop effect. This result confirms that the original finding is not sensitive to the centre tested. As additional benchmarks, we ran the maximum mean discrepancy (MMD) test of \textcite{Gretton2012} and the Bayesian Pólya tree test of \textcite{Holmes2015}, both tests for equality of distributions. The $p$-value and null probability is 0.499 and 0.223, respectively. These tests assume the samples are independent, so they are likely underpowered here.

\subsection{Multi-task perception study}

\textcite{Srna2018} ran an experiment to investigate if human performance at certain activities is affected by whether the activity is perceived as multi-tasking (refer to Study 1a in that paper). Experimental subjects were required to watch a video and transcribe the audio. This activity was framed as multi-tasking to a treatment group and single-tasking to a control group. Assignment to either group was random. The study found that subjects in the treatment group transcribed more words than those in the control group and the accuracy of their transcriptions was higher, suggesting perceiving an activity as multi-tasking improves performance at that activity.

\begin{figure}[ht!]
\centering
\includegraphics{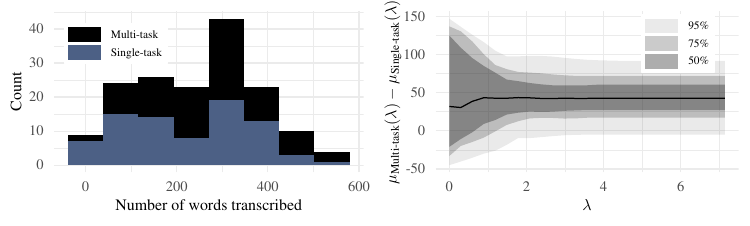}
\caption{Multi-task perception data. The left plot is a histogram of the data by control and treatment. The right plot is a functional boxplot of the posterior difference in Huber families. Shading indicates different central regions of the posterior.}
\label{fig:multitask-perception}
\end{figure}

We focus on the number of words transcribed. The dataset \parencite{Srna2018b} contains $n_1=82$ subjects in the treatment group and $n_2=80$ in the control group; see Figure~\ref{fig:multitask-perception}.
The groups are dissimilar in distribution, with the multi-task group being unimodal and the single-task group being multimodal. For small values of $\lambda$, the 50\% central region of the posterior includes zero, indicating the null might be plausible.

The study reported a $p$-value of $0.033$ from an $F$ test of the multi-task condition in a one-way ANOVA, identical to a two-sample Student $t$ test with equal variance that the mean number of words transcribed is equal between groups. The Mood median test yields a $p$-value of 0.271. The $p$-value from a Wilcoxon rank-sum test is 0.072. The MMD and Pólya tree tests produce a $p$-value and null probability of 0.590 and 0.978, respectively. The Huber familial test returns 0.170 as the probability of the null. In contrast to the $t$ test, the familial, sign, rank-sum, MMD, and Pólya tree tests do not find the multi-task condition to affect performance. In particular, the familial test fails to find sufficient support for either hypothesis and returns an indeterminate result. Whether this is a meaningful discrepancy remains up to subject-matter experts to decide.

\subsection{Simulations}

Appendix~\ref{app:simulations} contains extensive results on synthetic datasets that verify the finite-sample properties of the familial test in one- and two-sample settings. The results cover a variety of distributions, both continuous and discrete, and validate that the familial test is well-behaved. The test has good frequentist size and, for a sufficiently large departure from the null or a reasonably large sample size, high power. In fact, its power remains highly competitive with existing frequentist and Bayesian tests in regions where the familial null fails to hold.

\section{Concluding remarks}
\label{sec:conclusion}

It has become standard practice to translate scientific hypotheses into statistical hypotheses about a specific centre for the underlying distribution(s). Despite the ubiquity of this approach, there can be a lack of consensus about which centre bests reflects the original scientific hypotheses. When there is ambiguity, we argue one should adopt familial inference, which formulates hypotheses via a family of plausible centres. Our package \texttt{familial} implements the tools developed in this paper and is publicly available on \texttt{CRAN}.

A natural next step in this line of work is to develop familial inference for other statistical parameters and models. For instance, we found in ongoing work that our tools extend gracefully to linear models. The Huber family of linear models constitutes a continuum of models for conditional centres, from the conditional mean to the conditional median. The pathwise algorithm extends to this setting because the solution path remains piecewise linear under Huber loss.

Another intriguing direction is familial inference for multivariate random variables. The univariate Huber function $\ell_\lambda(z)$ studied in this paper has a lesser-known multivariate counterpart \parencite{Hampel1986}, which composes the $\ell_2$-norm with the univariate Huber function, i.e., $\ell_\lambda(\|\mathbf{z}\|)$. It remains to be determined whether the multivariate problem remains computationally tractable.

It is also interesting to consider other loss functions for the univariate, multivariate, and linear model problems. For instance, the univariate trimmed square loss (i.e., the trimmed mean), is also piecewise linear, and readily admits a pathwise algorithm. More generally, fundamentally different algorithms are required.

A frequentist version of the familial test is likewise an appealing avenue of future research. A major challenge is deriving the finite-sample, or asymptotic, distribution of the Huber family under the null hypothesis. A bootstrap test is also possible, but it remains to determine the appropriate way of imposing the null hypothesis on the observed sample to attain the correct size.

\section*{Acknowledgements}

Thompson acknowledges financial support by an Australian Government Research Training Program (RTP) Scholarship. Forbes, MacEachern, and Peruggia acknowledge financial support by the National Science Foundation (NSF) Grant SES-1921523. MacEachern also acknowledges financial support by the NSF Grant DMS-2015552.

\printbibliography

\begin{appendices}

\section{Huber family}
\label{app:huber}

\subsection{Proof of Lemma~\ref{lemma:crossing}}

\begin{proof}
A sufficient condition for the result of the lemma is for $\lambda-|x_0-\mu(\lambda)|$ to be increasing as a function of $\lambda$. To establish the function is increasing, consider its gradient:
\begin{equation}
\label{eq:crossing1}
\frac{\partial}{\partial\lambda}\left(\lambda-|x_0-\mu(\lambda)|\right)=1+\sign(x_0-\mu(\lambda))\frac{\partial\mu(\lambda)}{\partial\lambda}.
\end{equation}
A sufficient condition for the gradient to be positive, and hence for $\lambda-|x_0-\mu(\lambda)|$ to be increasing, is that $|\partial\mu(\lambda)/\partial\lambda|<1$. The first-order condition for optimality of $\mu(\lambda)$ is
\begin{equation*}
\mathcal{L}'(\mu(\lambda))=-\sum_{i:|x_i-\mu(\lambda)|<\lambda}w_i(x_i-\mu(\lambda))-\sum_{i:|x_i-\mu(\lambda)|\geq\lambda}w_i\lambda\sign(x_i-\mu(\lambda))=0,
\end{equation*}
which gives
\begin{equation}
\label{eq:crossing2}
-\frac{\sum_{i:|x_i-\mu(\lambda)|\geq\lambda}w_i\lambda\sign(x_i-\mu(\lambda))}{\sum_{i:|x_i-\mu(\lambda)|<\lambda}w_i(x_i-\mu(\lambda))}=1.
\end{equation}
Using the bound $|x_i-\mu(\lambda)|<\lambda$ in the denominator of \eqref{eq:crossing2} yields
\begin{equation*}
\begin{split}
-\frac{\sum_{i:|x_i-\mu(\lambda)|\geq\lambda}w_i\lambda\sign(x_i-\mu(\lambda))}{\sum_{i:|x_i-\mu(\lambda)|<\lambda}w_i(x_i-\mu(\lambda))}&>\left|-\frac{\sum_{i:|x_i-\mu(\lambda)|\geq\lambda}w_i\lambda\sign(x_i-\mu(\lambda))}{\sum_{i:|x_i-\mu(\lambda)|<\lambda}w_i\lambda}\right| \\
&=\left|-\frac{\sum_{i:|x_i-\mu(\lambda)|\geq\lambda}w_i\sign(x_i-\mu(\lambda))}{\sum_{i:|x_i-\mu(\lambda)|<\lambda}w_i}\right| \\
&=\left|\frac{\partial\mu(\lambda)}{\partial\lambda}\right|.
\end{split}
\end{equation*}
Together with \eqref{eq:crossing2}, the above bound shows $|\partial\mu(\lambda)/\partial\lambda|<1$, and hence the gradient~\eqref{eq:crossing1} is positive. We conclude $\lambda-|x_0-\mu(\lambda)|$ is increasing, thereby establishing the result of the lemma.
\end{proof}

\subsection{Proof of Proposition~\ref{prop:step}}

The proof of Proposition~\ref{prop:step} requires the following lemma.

\begin{lemma}
\label{lemma:sign}
Let $s_i:=\sign(x_i-\tilde{\mu})$, where $\tilde{\mu}$ is the weighted median. Suppose sample point $x_0$ satisfies $|x_0-\mu(\lambda^\star)|\geq\lambda^\star$ for some $\lambda^\star>0$. Then $\sign(x_0-\mu(\lambda^\star))=s_0$.
\begin{proof}
We proceed using proof by contradiction and suppose $\sign(x_0-\mu(\lambda^\star))\neq s_0$. This event can only occur if there exists a $0<\lambda<\lambda^\star$ such that $|x_0-\mu(\lambda)|<\lambda$, since for the sign of $x_0-\mu(\lambda)$ to change the residual must cross through zero. But the existence of such a $\lambda$ contradicts Lemma~\ref{lemma:crossing} since $|x_0-\mu(\lambda^\star)|\geq\lambda^\star$. Hence, it must be the case that $\sign(x_0-\mu(\lambda^\star))=\sign(x_0-\mu(\lambda))$ for all $0<\lambda<\lambda^\star$. The result of the lemma immediately follows from the fact that $\lim_{\lambda\to0}\mu(\lambda)=\tilde{\mu}$.
\end{proof}
\end{lemma}

We are now ready to prove Proposition~\ref{prop:step}.

\begin{proof}
By equation~\eqref{eq:lambda+}, $\gamma=\lambda-\lambda^+$. Since $(\lambda^+,\mu^+)$ is a knot point, one or more sample points cross from the square piece of the Huber function to the absolute piece and satisfy $|x_i-\mu^+|=\lambda^+$. Among all sample points eligible to cross (i.e., all $i$ satisfying $|x_i-\mu|<\lambda$), those with with maximal absolute deviation from $\mu^+$ cross:
\begin{equation*}
\lambda^+=\underset{i:|x_i-\mu|<\lambda}{\max}\left(|x_i-\mu^+|\right).
\end{equation*}
Together, the above expressions for $\gamma$ and $\lambda^+$ give
\begin{equation*}
\gamma=\lambda-\underset{i:|x_i-\mu|<\lambda}{\max}\left(|x_i-\mu^+|\right)=\min_{i:|x_i-\mu|<\lambda}\left(\lambda-|x_i-\mu^+|\right).
\end{equation*}
Since $|x_i-\mu^+|=\lambda^+$ for $i$ satisfying the above equalities, we can invoke Lemma~\ref{lemma:sign} to get
\begin{equation*}
\gamma=\min_{i:|x_i-\mu|<\lambda}\left(\lambda-s_i(x_i-\mu^+)\right).
\end{equation*}
Now, making the substitution $\mu^+=\mu-\gamma\partial\mu(\lambda)/\partial\lambda$ per equation~\eqref{eq:mu+} and rearranging terms leads to
\begin{equation}
\label{eq:step1}
0=\min_{i:|x_i-\mu|<\lambda}\left(\lambda-s_i(x_i-\mu)-\gamma\left(1+s_i\frac{\partial\mu(\lambda)}{\partial\lambda}\right)\right).
\end{equation}
We have $1+s_i\partial\mu(\lambda)/\partial\lambda>0$ since $|\partial\mu(\lambda)/\partial\lambda|<1$, as established in the proof of Lemma~\ref{lemma:crossing}. Hence, equality~\eqref{eq:step1} remains valid after division by $1+s_i\partial\mu(\lambda)/\partial\lambda$ inside the minimization. Performing the division and isolating $\gamma$ yields
\begin{equation*}
\gamma=\min_{i:|x_i-\mu|<\lambda}\left(\frac{\lambda-s_i(x_i-\mu)}{1+s_i\partial\mu(\lambda)/\partial\lambda}\right),
\end{equation*}
as per the result of the proposition.
\end{proof}

\section{Consistency}
\label{app:consistency}

\subsection{Proof of Theorem~\ref{thm:consistency}}

We begin by introducing some notation. Denote the empirical distribution function by
\begin{equation*}
P_n(x):=\sum_{i=1}^n\frac{1}{n}1(X_i\leq x)
\end{equation*}
and the random Bayesian bootstrap distribution function by
\begin{equation*}
G_n(x):=\sum_{i=1}^nw_i1(X_i\leq x).
\end{equation*}
Denote the integral of the gradient of the Huber function under some distribution $P$ by
\begin{equation*}
A(P,\lambda,\mu):=\int\ell_\lambda'(X_i-\mu)\,dP(x).
\end{equation*}
The gradient $\ell_\lambda'(x-\mu)$ is Lipschitz continuous in $\lambda$, $x$, and $\mu$ with Lipschitz constant 1.

Our work focuses on the gradient of the Huber function, since its roots define the Huber family. The proof of the theorem requires several technical lemmas, which we now state and prove before arriving at the main argument.

We begin with Lemma~\ref{lemma:bound_one_lambda}, which provides a probabilistic finite-sample bound for the distance between $A(G_n,\lambda,\mu)$---the integral under a Bayesian bootstrap distribution---and $A(P_n,\lambda,\mu)$---the integral under the empirical distribution. This bound holds for fixed $\lambda\in\Lambda$.

\begin{lemma}
\label{lemma:bound_one_lambda}
Let $X_1,\ldots, X_n$ be iid random variables with distribution $P_0$ and, independently, let $(w_1,\ldots,w_n)\sim\dir(1,\ldots,1)$. Let $\epsilon>0$ and $\lambda\geq0$. Then, for any $\mu\in\mathbb{R}$, it holds
\begin{equation*}
\p(|A(G_n,\lambda,\mu)-A(P_n,\lambda,\mu)|\geq\epsilon)\leq\frac{\lambda^2}{\epsilon^2}\,\frac{n-1}{n+1}\,\frac{1}{n}.
\end{equation*}
\end{lemma}

\begin{proof}
Since $(w_1,\ldots,w_n)$ are $\dir(1,...,1)$, we have
\begin{equation*}
\begin{split}
\e(w_i)&=\frac{1}{n}, \\
\var(w_i)&=\frac{1}{n+1}\frac{1}{n}\frac{n-1}{n},\text{ and} \\
\cov(w_i,w_j)&=-\frac{1}{n+1}\frac{1}{n^2}\text{ for }i\neq j.
\end{split}
\end{equation*}
First, observe that
\begin{equation*}
\begin{split}
\e\left[A(G_n,\lambda,\mu)-A(P_n,\lambda,\mu)\right]&=\sum_{i=1}^n\e\left[\left(w_i-\frac{1}{n}\right)\ell_\lambda'(X_i-\mu)\right] \\
&=\sum_{i=1}^n\e\left\{\e\left[\left(w_i-\frac{1}{n}\right)\ell_\lambda'(X_i-\mu)\right]\,\middle|\, w_i\right\} \\
&=\sum_{i=1}^n\e\left[\left(w_i-\frac{1}{n}\right)\right]\e\left[\ell_\lambda'(X_1-\mu)\right] \\
&=0.
\end{split}
\end{equation*}
For the variance, we have
\begin{equation}
\label{eq:var}
\begin{split}
&\var\left[A(G_n,\lambda,\mu)-A(P_n,\lambda,\mu)\right] \\
&\hspace{2cm}=\var\left[\sum_{i=1}^n\left(w_i-\frac{1}{n}\right)\ell_\lambda'(X_i-\mu)\right] \\
&\hspace{2cm}=\var\left\{\e\left[\sum_{i=1}^n\left(w_i-\frac{1}{n}\right)\ell_\lambda'(X_i-\mu)\,\middle|\,(w_1,\ldots,w_n)\right]\right\} \\
&\hspace{3cm}+\e\left\{\var\left[\sum_{i=1}^n\left(w_i-\frac{1}{n}\right)\ell_\lambda'(X_i-\mu)\,\middle|\,(w_1,\ldots,w_n)\right]\right\}.
\end{split}
\end{equation}
The first-term on the right-hand side of \eqref{eq:var} is zero, since
\begin{equation*}
\begin{split}
&\var\left\{\e\left[\sum_{i=1}^n\left(w_i-\frac{1}{n}\right)\ell_\lambda'(X_i-\mu)\,\middle|\,(w_1,\ldots,w_n)\right]\right\} \\
&\hspace{5cm}=\var\left\{\sum_{i=1}^n\left(w_i-\frac{1}{n}\right)\e\left[\ell_\lambda'(X_i-\mu)\right]\right\} \\
&\hspace{5cm}=\var(0)=0.
\end{split}
\end{equation*}
The second term on the right-hand side of \eqref{eq:var} satisfies
\begin{equation*}
\begin{split}
&\e\left\{\var\left[\sum_{i=1}^n\left(w_i-\frac{1}{n}\right)\ell_\lambda'(X_i-\mu)\,\middle|\,(w_1,\ldots,w_n)\right]\right\} \\
&\hspace{5cm}=\e\left\{\var\left[\ell_\lambda'(X_1-\mu)\right]\sum_{i=1}^n\left(w_i-\frac{1}{n}\right)^2\right\} \\
&\hspace{5cm}=\var\left[\ell_\lambda'(X_1-\mu)\right]n\frac{1}{n+1}\frac{1}{n}\frac{n-1}{n} \\
&\hspace{5cm}\leq\lambda^2\frac{n-1}{n+1}\frac{1}{n},
\end{split}
\end{equation*}
where the inequality follows from the fact that, for all $\mu\in\mathbb{R}$, the gradient $\ell_\lambda'(X_i-\mu)$ is bounded between $-\lambda$ and $\lambda$. The result of the lemma now follows directly from Chebyshev's inequality.
\end{proof}

Lemma~\ref{lemma:bound_one_lambda} holds for fixed $\lambda\in\Lambda$. We now state and prove Lemma~\ref{lemma:bound_all_lambda}, which extends the bound in Lemma~\ref{lemma:bound_one_lambda} to hold uniformly for all $\lambda\in\Lambda$.

\begin{lemma}
\label{lemma:bound_all_lambda}
Let $X_1,\ldots,X_n$ be iid random variables with distribution $P_0$ and, independently, let $(w_1,\ldots,w_n)\sim\dir(1,\ldots,1)$. Let $\epsilon>0$ and $\Lambda=[0,\bar{\lambda}]$ with $0<\bar{\lambda}<\infty$. Then, for any $\mu\in\mathbb{R}$, it holds
\begin{equation*}
\p\left(\sup_{\lambda\in\Lambda}\left|A(G_n,\lambda,\mu)-A(P_n,\lambda,\mu)\right|\geq\epsilon\right)\leq\left(\left\lceil\frac{4\bar{\lambda}}{\epsilon}\right\rceil+1\right)\frac{4\bar{\lambda}^2}{\epsilon^2}\frac{n-1}{n+1}\frac{1}{n}.
\end{equation*}
\end{lemma}

\begin{proof}
We proceed using an epsilon-net argument. First, we require a useful inequality. For any $\lambda\neq\lambda'$, we have
\begin{equation*}
\begin{split}
\left|A(G_n,\lambda,\mu)-A(P_n,\lambda,\mu)\right|&=\left|A(G_n,\lambda,\mu)-A(G_n,\lambda',\mu)\right. \\
&\hspace{1cm}+A(G_n,\lambda',\mu)-A(P_n,\lambda',\mu)+ \\
&\hspace{2cm}\left.A(P_n,\lambda',\mu)-A(P_n,\lambda,\mu)\right| \\
&\leq\left|A(G_n,\lambda,\mu)-A(G_n,\lambda',\mu)\right| \\
&\hspace{1cm}+\left|A(G_n,\lambda',\mu)-A(P_n,\lambda',\mu)\right| \\
&\hspace{2cm}+\left|A(P_n,\lambda',\mu)-A(P_n,\lambda,\mu)\right| \\
&\leq |\lambda-\lambda'|+\left|A(G_n,\lambda',\mu)-A(P_n,\lambda',\mu)\right|+|\lambda'-\lambda| \\
&=\left|A(G_n,\lambda',\mu)-A(P_n,\lambda',\mu)\right|+2|\lambda-\lambda'|.
\end{split}
\end{equation*}
The last inequality follows from the fact that $\ell_\lambda'(X-\mu)$ is Lipschitz in $\lambda$ with constant 1 (also, at $\lambda=0$ the left-hand side will be zero). Now, let $\mathcal{E}$ be an $\delta$-net of $\Lambda$. For any $\lambda\in\Lambda$, there exists a $\lambda'\in\mathcal{E}$ such that $|\lambda-\lambda'|\leq\delta$. This result in combination with the previous bound gives
\begin{equation*}
\left|A(G_n,\lambda,\mu)-A(P_n,\lambda,\mu)\right|\leq\left|A(G_n,\lambda',\mu)-A(P_n,\lambda',\mu)\right|+2\delta.
\end{equation*}
Taking the maximum of the right-hand side and the supremum of the left-hand side yields
\begin{equation*}
\sup_{\lambda\in\Lambda}\left|A(G_n,\lambda,\mu)-A(P_n,\lambda,\mu)\right|\leq\max_{\lambda'\in\mathcal{E}}\left|A(G_n,\lambda',\mu)-A(P_n,\lambda',\mu)\right|+2\delta.
\end{equation*}
It then follows
\begin{equation*}
\begin{split}
&\p\left(\sup_{\lambda\in\Lambda}\left|A(G_n,\lambda,\mu)-A(P_n,\lambda,\mu)\right|\geq\epsilon\right) \\
&\hspace{4cm}\leq\p\left(\max_{\lambda'\in\mathcal{E}}\left|A(G_n,\lambda',\mu)-A(P_n,\lambda',\mu)\right|+2\delta\geq\epsilon\right).
\end{split}
\end{equation*}
Next, setting $\delta=\epsilon/4$ and applying a union bound on the right-hand side, gives
\begin{equation*}
\begin{split}
&\p\left(\max_{\lambda'\in\mathcal{E}}\left|A(G_n,\lambda',\mu)-A(P_n,\lambda',\mu)\right|\geq\frac{1}{2}\epsilon\right) \\
&\hspace{4cm}\leq\sum_{\lambda'\in\mathcal{E}}\p\left(\left|A(G_n,\lambda',\mu)-A(P_n,\lambda',\mu)\right|\geq\frac{1}{2}\epsilon\right).
\end{split}
\end{equation*}
We can bound elements of the sum by Lemma~\ref{lemma:bound_one_lambda} as
\begin{equation*}
\p\left(\left|A(G_n,\lambda',\mu)-A(P_n,\lambda',\mu)\right|\geq\frac{1}{2}\epsilon\right)\leq\frac{4\lambda'^2}{\epsilon^2}\frac{n-1}{n+1}\frac{1}{n}\leq\frac{4\bar{\lambda}^2}{\epsilon^2}\frac{n-1}{n+1}\frac{1}{n}.
\end{equation*}
The set $\mathcal{E}$ has cardinality at most $\lceil\bar{\lambda}/\delta\rceil+1$, so
\begin{equation*}
\sum_{\lambda'\in\mathcal{E}}\p\left(\left|A(G_n,\lambda',\mu)-A(P_n,\lambda',\mu)\right|\geq\frac{1}{2}\epsilon\right)\leq\left(\left\lceil\frac{4\bar{\lambda}}{\epsilon}\right\rceil+1\right)\frac{4\bar{\lambda}^2}{\epsilon^2}\frac{n-1}{n+1}\frac{1}{n}.
\end{equation*}
The statement of the theorem now follows.
\end{proof}

We now turn to Lemma~\ref{lemma:neighbourhood}. Recall the Levy distance between two distributions $G$ and $F$ is defined as
\begin{equation*}
d_L(G,F):=\inf\{\epsilon\mid F(x-\epsilon)-\epsilon\leq G(x)\leq F(x+\epsilon)+\epsilon\text{ for all } x\in\mathbb{R}\}.
\end{equation*}
The following lemma provides a non-probabilistic bound on the distance between $A(G,\lambda,\mu)$ and $A(F,\lambda,\mu)$ when $G$ and $F$ are in an $\epsilon$-neighbourhood under the Levy metric.

\begin{lemma}
\label{lemma:neighbourhood}
Let $F$ be a distribution and $\epsilon>0$. Let $\Lambda=[0,\bar{\lambda}]$ for some $0<\bar{\lambda}<\infty$. Consider an open neighbourhood of $F$, given by $U=\{G: d_L(F,G)<\epsilon/2\}$. Then, for all $G\in U$, it holds
\begin{equation*}
\sup_{\lambda\in\Lambda}|A(G,\lambda,\mu)-A(F,\lambda,\mu)|<\epsilon.
\end{equation*}
\end{lemma}

\begin{proof}
We begin with a fixed $\lambda\in\Lambda$ and then extend to a uniform bound for all $\lambda\in\Lambda$. Let $\delta=\epsilon/2$. Then, for any $\lambda$, we have
\begin{equation}
\label{eq:neighbourhood_one_lambda}
\begin{split}
\left|A(G,\lambda,\mu)-A(F,\lambda,\mu)\right|&=\left|\int\ell_\lambda'(x-\mu)\,dG(x)-\int\ell_\lambda'(x-\mu)\,dF(x+\delta)\right. \\
&\hspace{1cm}+\left.\int\ell_\lambda'(x-\mu)\,dF(x+\delta)-\int\ell_\lambda'(x-\mu)\,dF(x)\right| \\
&\leq\left|\delta+\int\ell_\lambda'(x-\mu)\,dF(x+\delta)~-\int\ell_\lambda'(x-\mu)\,dF(x)\right|	\\
&\leq\delta+\int\left|\ell_\lambda'(x-\delta-\mu)-\ell_\lambda'(x-\mu)\right|\,dF(x)\\
&\leq\delta+\delta=2\delta.
\end{split}
\end{equation}
The first inequality is due to $d_L(G,F)<\delta$ and the third due to $\ell_\lambda'(x-\mu)$ being Lipschitz in $x$ with constant 1. We now extend this result to hold over all $\lambda\in\Lambda$ by taking the supremum of the left-hand side of \eqref{eq:neighbourhood_one_lambda}, yielding
\begin{equation*}
\sup_{\lambda\in\Lambda}|A(G,\lambda,\mu)-A(F,\lambda,\mu)|\leq2\delta.
\end{equation*}
The statement of the lemma now follows from the fact that $\delta=\epsilon/2$.
\end{proof}

Lemma~\ref{lemma:converge} is the last piece required before establishing the result of the theorem. It draws on the preceding lemmas to prove that $A(G_n,\lambda,\mu)$ converges in probability to $A(P_0,\lambda,\mu)$ uniformly for all $\lambda\in\Lambda$.

\begin{lemma}
\label{lemma:converge}
Let $X_1,\ldots,X_n$ be iid random variables with distribution $P_0$ and, independently, let $(w_1,\ldots,w_n)\sim\dir(1,\ldots,1)$. Let $\epsilon>0$ and $\Lambda=[0,\bar{\lambda}]$ with $0<\bar{\lambda}<\infty$. Then, for any $\mu\in\mathbb{R}$, it holds
\begin{equation*}
\lim_{n\to\infty}\p\left(\sup_{\lambda\in\Lambda}\left|A(G_n,\lambda,\mu)-A(P_0,\lambda,\mu)\right|\geq\epsilon\right)=0.
\end{equation*}
\end{lemma}

\begin{proof}
We begin with the inequality
\begin{equation*}
\begin{split}
|A(G_n,\lambda,\mu)-A(P_0,\lambda,\mu)|&=|A(G_n,\lambda,\mu)-A(P_n,\lambda,\mu)+ \\
&\hspace{2cm} A(P_n,\lambda,\mu)-A(P_0,\lambda,\mu)| \\
&\leq|A(G_n,\lambda,\mu)-A(P_n,\lambda,\mu)|+ \\
&\hspace{2cm}|A(P_n,\lambda,\mu)-A(P_0,\lambda,\mu)|.
\end{split}
\end{equation*}
Taking the supremum over both sides gives
\begin{equation*}
\begin{split}
\sup_{\lambda\in\Lambda}|A(G_n,\lambda,\mu)-A(P_0,\lambda,\mu)|&\leq\sup_{\lambda\in\Lambda}|A(G_n,\lambda,\mu)-A(P_n,\lambda,\mu)|+ \\
&\hspace{2cm}\sup_{\lambda\in\Lambda}|A(P_n,\lambda,\mu)-A(P_0,\lambda,\mu)|.
\end{split}
\end{equation*}
It follows
\begin{equation}
\label{eq:bound}
\begin{split}
&\p\left(\sup_{\lambda\in\Lambda}|A(G_n,\lambda,\mu)-A(P_0,\lambda,\mu)|\geq\epsilon\right) \\
&\hspace{3cm}\leq\p\left(\sup_{\lambda\in\Lambda}|A(G_n,\lambda,\mu)-A(P_n,\lambda,\mu)|\geq\epsilon\right)+ \\
&\hspace{5cm}\p\left(\sup_{\lambda\in\Lambda}|A(P_n,\lambda,\mu)-A(P_0,\lambda,\mu)|\geq\epsilon\right).
\end{split}
\end{equation}
We now bound terms on the right-hand side of \eqref{eq:bound}. For the first term, we can apply Lemma~\ref{lemma:bound_all_lambda}. Given $\epsilon>0$, there exists an $N_1$ such that for all $n\geq N_1$, it holds
\begin{equation*}
\p\left(\sup_{\lambda\in\Lambda}|A(G_n,\lambda,\mu)-A(P_n,\lambda,\mu)|\geq\epsilon\right)\leq\frac{\epsilon}{2}.
\end{equation*}
Consider the second term on the right-hand side of \eqref{eq:bound}. We have that $P_n$ converges almost surely to $P_0$ by the Glivenko-Cantelli theorem and hence to a $\delta$-neighbourhood of $P_0$ under the Levy metric. Set $\delta=\epsilon/2$. This result and Lemma~\ref{lemma:neighbourhood} provide there exists an $N_2$ such that for all $n\geq N_2$, it holds
\begin{equation*}
\p\left(\sup_{\lambda\in\Lambda}|A(P_n,\lambda,\mu)-A(P_0,\lambda,\mu)|\geq\epsilon\right)\leq\frac{\epsilon}{2}.
\end{equation*}
Set $N=\max(N_1,N_2)$. Then, for all $n\geq N$, we can bound the left-hand side of \eqref{eq:bound} as
\begin{equation*}
\p\left(\sup_{\lambda\in\Lambda}|A(G_n,\lambda,\mu)-A(P_0,\lambda,\mu)|\geq\epsilon\right)\leq\epsilon
\end{equation*}
The result of the lemma now follows.
\end{proof}

We now have everything necessary to prove the two claims of Theorem~\ref{thm:consistency}.

\begin{proof}

We consider case 1 ($\mathrm{H}_0$ is true) and case 2 ($\mathrm{H}_1$ is true) in turn. To simplify the proof's presentation, we assume the hypothesized null value $m_0=0$ without loss of generality.

\subsubsection*{Case 1: $\mathrm{H}_0$ is true}

Suppose the null hypothesis is true and $\mu_0(\lambda)$ strictly crosses through zero. In this case, it must hold that $A(P_0,\lambda,0)>c>0$ for some $\lambda\in\Lambda$ and $A(P_0,\lambda,0)<-c<0$ for some $\lambda\in\Lambda$. It follows that both
\begin{equation*}
\sup_{\lambda\in\Lambda}A(P_0,\lambda,0)\geq c>0
\end{equation*}
and
\begin{equation*}
\inf_{\lambda\in\Lambda}A(P_0,\lambda,0)\leq-c<0
\end{equation*}
hold simultaneously. Our task is to show that $\sup_{\lambda\in\Lambda}A(G_n,\lambda,0)$ is bounded below away from zero and $\inf_{\lambda\in\Lambda}A(G_n,\lambda,0)$ is bounded above away from zero, respectively, with probability tending to one. For the supremum, we have
\begin{equation*}
\begin{split}
\sup_{\lambda\in\Lambda}A(G_n,\lambda,0)&=\sup_{\lambda\in\Lambda}(A(P_0,\lambda,0)+A(G_n,\lambda,0)-A(P_0,\lambda,0)) \\
&\geq\sup_{\lambda\in\Lambda}(A(P_0,\lambda,0)-|A(G_n,\lambda,0)-A(P_0,\lambda,0)|) \\
&\geq\sup_{\lambda\in\Lambda}A(P_0,\lambda,0)-\sup_{\lambda\in\Lambda}|A(G_n,\lambda,0)-A(P_0,\lambda,0)|.
\end{split}
\end{equation*}
Likewise, for the infimum, it holds
\begin{equation*}
\begin{split}
\inf_{\lambda\in\Lambda}A(G_n,\lambda,0)&=\inf_{\lambda\in\Lambda}(A(P_0,\lambda,0)+A(G_n,\lambda,0)-A(P_0,\lambda,0)) \\
&\leq\inf_{\lambda\in\Lambda}(A(P_0,\lambda,0)+|A(G_n,\lambda,0)-A(P_0,\lambda,0)|) \\
&\leq\inf_{\lambda\in\Lambda}A(P_0,\lambda,0)+\sup_{\lambda\in\Lambda}|A(G_n,\lambda,0)-A(P_0,\lambda,0)|.
\end{split}
\end{equation*}
Now, fix any $0<\epsilon<c$. Then, by Lemma~\ref{lemma:converge}, there exists an $N$ such that for all $n\geq N$, it holds with probability at least $1-\epsilon$ that
\begin{equation*}
\begin{split}
\sup_{\lambda\in\Lambda}A(G_n,\lambda,0)&\geq\sup_{\lambda\in\Lambda}A(P_0,\lambda,0)-\epsilon \\
&\geq c-\epsilon>0
\end{split}
\end{equation*}
and
\begin{equation*}
\begin{split}
\inf_{\lambda\in\Lambda}A(G_n,\lambda,0)&\leq\inf_{\lambda\in\Lambda}A(P_0,\lambda,0)+\epsilon \\
&\leq -c+\epsilon<0.
\end{split}
\end{equation*}
Hence, the supremum is bounded below by zero and the infimum is bounded above by zero, with high probability. Thus, with probability tending to one, it must hold that $A(G_n,\lambda,0)=0$ for some $\lambda\in\Lambda$, and hence (by continuity of the solution path) that $\hat{\mu}(\lambda)=0$ for some $\lambda\in\Lambda$. This result concludes the first claim of the theorem.

\subsubsection*{Case 2: $\mathrm{H}_1$ is true}

Suppose the alternative hypothesis is true and $\mu_0(\lambda)$ is bounded away from zero for all $\lambda\in\Lambda$. In this case, it must hold that (a) $A(P_0,\lambda,0)\leq-c_1<0$ for all $\lambda\in\Lambda$ or (b) $A(P_0,\lambda,0)\geq c_2>0$ for all $\lambda\in\Lambda$. For (a), it follows that 
\begin{equation*}
\sup_{\lambda\in\Lambda}A(P_0,\lambda,0)\leq-c_1<0,
\end{equation*}
while (b) corresponds to
\begin{equation*}
\inf_{\lambda\in\Lambda}A(P_0,\lambda,0)\geq c_2>0.
\end{equation*}
Our task for (a) is to show that $\sup_{\lambda\in\Lambda}A(G_n,\lambda,0)$ is also bounded above away from zero, and for (b) that $\inf_{\lambda\in\Lambda}A(G_n,\lambda,0)$ is also bounded below away from zero, in both situations with probability tending to one. Consider (a) first. We have
\begin{equation*}
\begin{split}
\sup_{\lambda\in\Lambda}A(G_n,\lambda,0)&=\sup_{\lambda\in\Lambda}(A(P_0,\lambda,0)+A(G_n,\lambda,0)-A(P_0,\lambda,0)) \\
&\leq\sup_{\lambda\in\Lambda}(A(P_0,\lambda,0)+|A(G_n,\lambda,0)-A(P_0,\lambda,0)|) \\
&\leq\sup_{\lambda\in\Lambda}A(P_0,\lambda,0)+\sup_{\lambda\in\Lambda}|A(G_n,\lambda,0)-A(P_0,\lambda,0)|.
\end{split}
\end{equation*}
Fix any $0<\epsilon_1<c_1$. By Lemma~\ref{lemma:converge}, there exists an $N_1$ such that for all $n\geq N_1$, it holds with probability at least $1-\epsilon_1$ that
\begin{equation*}
\begin{split}
\sup_{\lambda\in\Lambda}A(G_n,\lambda,0)&\leq\sup_{\lambda\in\Lambda}A(P_0,\lambda,0)+\epsilon_1 \\
&\leq-c_1+\epsilon_1<0.
\end{split}
\end{equation*}
Hence, $\sup_{\lambda\in\Lambda}A(G_n,\lambda,0)$ is bounded above away from zero with high probability. Turning now to (b), we have
\begin{equation*}
\begin{split}
\inf_{\lambda\in\Lambda}A(G_n,\lambda,0)&=\inf_{\lambda\in\Lambda}(A(P_0,\lambda,0)+A(G_n,\lambda,0)-A(P_0,\lambda,0)) \\
&\geq\inf_{\lambda\in\Lambda}(A(P_0,\lambda,0)-|A(G_n,\lambda,0)-A(P_0,\lambda,0)|) \\
&\geq\inf_{\lambda\in\Lambda}A(P_0,\lambda,0)-\sup_{\lambda\in\Lambda}|A(G_n,\lambda,0)-A(P_0,\lambda,0)|.
\end{split}
\end{equation*}
Again, fix any $0<\epsilon_2<c_2$. Then there exists an $N_2$ such that for all $n\geq N_2$, it holds with probability at least $1-\epsilon_2$ that
\begin{equation*}
\begin{split}
\inf_{\lambda\in\Lambda}A(G_n,\lambda,0)&\geq\inf_{\lambda\in\Lambda}A(P_0,\lambda,0)-\epsilon_2 \\
&\geq c_2-\epsilon_2>0.
\end{split}
\end{equation*}
Hence, $\inf_{\lambda\in\Lambda}A(G_n,\lambda,0)$ is bounded below away from zero with high probability. Thus, with probability tending it one, it must hold that (a) $A(G_n,\lambda,0)>0$ for all $\lambda\in\Lambda$ and hence that $\hat{\mu}(\lambda)>0$ for all $\lambda\in\Lambda$ or (b) $A(G_n,\lambda,0)<0$ for all $\lambda\in\Lambda$ and hence that $\hat{\mu}(\lambda)<0$ for all $\lambda\in\Lambda$. This result concludes the second claim of the theorem.

\end{proof}

\section{Intersection-union testing}
\label{app:iutesting}

The familial test we propose may be considered to have an \emph{intersection-union} test format. Introduced by \textcite{Berger1982}, an intersection-union test for a parameter $\theta\in\Theta$ is a test involving a null hypothesis that is a union of sets and an alternative hypothesis that is an intersection of sets. Specifically, letting $\Theta_j$ denote a subset of $\Theta$ for $j=1,2,\ldots,k$, an intersection-union test evaluates the hypotheses
\begin{equation}
\label{eq:iuhyp}
\mathrm{H}_0:\theta\in\cup_{j=1}^k\Theta_j\quad\text{vs.}\quad\mathrm{H}_1:\theta\in\cap_{j=1}^k\Theta_j^c,
\end{equation}
where $\Theta_j^c$ is the complement of $\Theta_j$. If $\mathrm{H}_0$ is true, $\theta$ must be contained in at least one of the $\Theta_j$ subsets. Hence, to conduct an intersection-union test, it suffices to perform $k$ separate tests of
\begin{equation*}
\mathrm{H}_{0j}:\theta\in\Theta_j\quad\text{vs.}\quad\mathrm{H}_{1j}:\theta\in\Theta_j^c
\end{equation*}
and then reject the overall null hypothesis $\mathrm{H}_0$ if and only if all $k$ individual null hypotheses $\mathrm{H}_{0j}$ are rejected. \textcite{Berger1982} proves the overall type I error rate of this procedure is no bigger than $\alpha$ if the individual tests are conducted with level $\alpha$. \textcite{Berger1982} also states conditions under which intersection-union tests have size exactly equal to $\alpha$, since they are generally conservative with type I error rate less than $\alpha$. \textcite{Berger1996} generalize these conditions and also provide an example where an initially conservative intersection-union test can be modified to improve its frequentist power characteristics. \textcite{Li2020} and \textcite{Yin2021} contain some recent applications.

The connection between intersection-union tests and our familial test arises from the fact that the familial null and alternative can be broken down into a collection of individual hypotheses, each concerning a different centre indexed by $\lambda$:
\begin{equation*}
\mathrm{H}_{0\lambda}:\mu(\lambda)\in\mathcal{M}_0\quad\text{vs.}\quad\mathrm{H}_{1\lambda}:\mu(\lambda)\in\mathcal{M}_1.
\end{equation*}
Recall that $\mathcal{M}_0$ and $\mathcal{M}_1$ are a partition of the parameter space, so $\mathcal{M}_1=\mathcal{M}_0^c$. Similar to an intersection-union test, the familial test rejects if and only if the individual null hypotheses $\mathrm{H}_{0\lambda}$ are rejected for all $\lambda\in\Lambda$. Consequently, the overall hypotheses can be expressed using a union and intersection:
\begin{equation*}
\mathrm{H}_0:\cup_{\lambda\in\Lambda}\{\mu(\lambda)\in\mathcal{M}_0\}\quad\text{vs.}\quad\mathrm{H}_1:\cap_{\lambda\in\Lambda}\{\mu(\lambda)\in \mathcal{M}_1\}.
\end{equation*}
Here, the union and intersection are with respect to the individual events $\{\mu(\lambda)\in\mathcal{M}_0\}$ and $\{\mu(\lambda)\in\mathcal{M}_1\}$, rather than subsets of the parameter spaces as with the intersection-union test. Of course, the intersection-union hypotheses~\eqref{eq:iuhyp} can also be expressed in terms of individual events as $\mathrm{H}_0:\cup_{j=1}^k\{\theta\in\Theta_j\}\text{ vs. }\mathrm{H}_1:\cap_{j=1}^k\{\theta\in\Theta_j^c\}$. A key difference between the tests, however, is that the familial test involves an uncountable number of events, whereas the number of events $k$ is typically finite in an intersection-union test. Though the Bayesian nonparametric procedure outlined in Section~\ref{sec:test} does not formally control the size of the test, it is insightful to consider its size and power properties in repeated sampling experiments, an exercise undertaken next.

Intersection-union tests are the opposite of union-intersection tests \parencite{Roy1953}, where the null is an intersection of sets and the alternative is a union of sets. Though the familial hypotheses we consider do not conform to this format, the analogous null hypothesis would impose that all familial centres lie in the null set and the alternative that at least one centre does not. For the Huber family, these hypotheses would correspond to a test of symmetry when $\mathcal{M}_0$ is a singleton. Though tests of symmetry can address some distributional questions, they do not capture the nuances of differing centres, potentially leaving out scientifically relevant insights.

\section{Simulations}
\label{app:simulations}

\subsection{One-sample and paired samples}

\begin{figure}[ht!]
\centering
\includegraphics{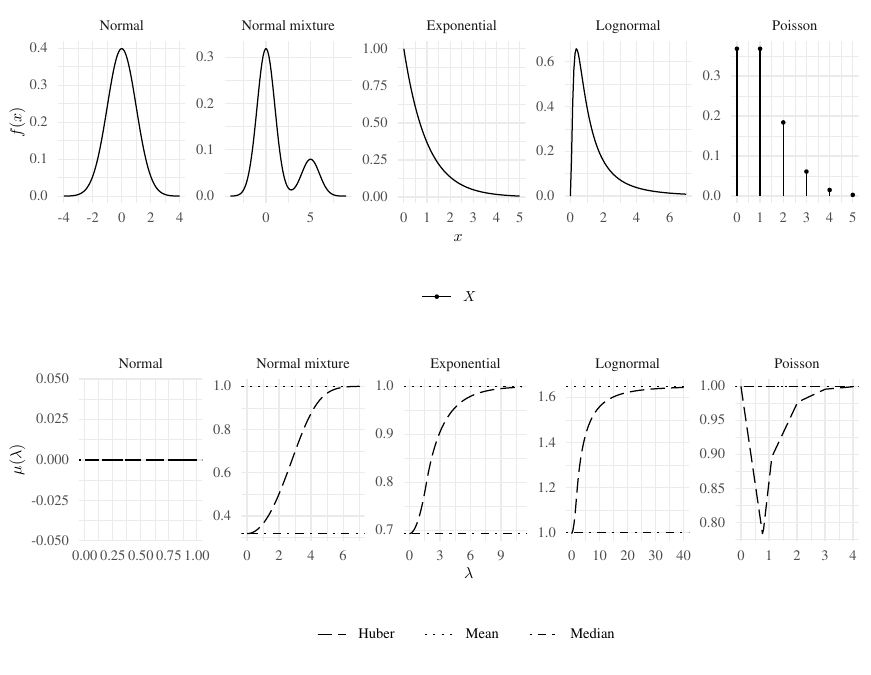}
\caption{Distributions analysed in the one-sample (paired samples) setting. The plots in the top row depict the density or mass function for the population. The plots in the bottom row depict the corresponding Huber family.}
\label{fig:one-sample-families}
\end{figure}

We first study the one-sample setting with $X_1,\ldots,X_n$. This setting can also be interpreted as the paired samples setting where $X_i$ is the difference of random variables. The distributions analysed are: $X\sim\mathrm{Normal}(0,1)$; $X\sim0.8\cdot\mathrm{Normal}(0,1)+0.2\cdot\mathrm{Normal}(5,1)$; $X\sim\mathrm{Exponential}(1)$; $X\sim\mathrm{Lognormal}(0,1)$; and $X\sim\mathrm{Poisson}(1)$. These distributions cover different types of support, modality, skewness, and tail behaviour. Figure~\ref{fig:one-sample-families} visualizes the distributions and their Huber families.
For the normal, the family is a singleton. For the normal mixture, exponential, and lognormal, the family is an interval with the mean and median as its endpoints. As the Poisson demonstrates, the family need not be bounded by the mean and median.

Table~\ref{tab:one-sample-tests} summarizes the tests evaluated and their associated hypotheses.
\begin{table}[ht!]
\centering
\small
\begin{threeparttable}
\begin{tabular}{llll}
\toprule
Test & Null hypothesis ($\mathrm{H}_0)$ & Alternative hypothesis ($\mathrm{H}_1)$ & Center \\
\midrule
Huber familial & $\exists\,\lambda\in\Lambda:\mu(\lambda)=m_0$ & $\forall\,\lambda\in\Lambda:\mu(\lambda)\neq m_0$ & Huber \\
Student $t$ & $\mu=m_0$ & $\mu\neq m_0$ & Mean \\
Fisher sign & $\mu=m_0$ & $\mu\neq m_0$ & Median \\
Wilcoxon signed-rank & $\mu=m_0$ & $\mu\neq m_0$ & Median\tnote{*} \\
\bottomrule
\end{tabular}
\begin{tablenotes}\footnotesize
\item [*] Provided $X$ is symmetric
\end{tablenotes}
\end{threeparttable}
\caption{Tests evaluated in the one-sample (paired samples) setting.}
\label{tab:one-sample-tests}
\end{table}
A Bayesian adaptation of the signed-rank test developed by \textcite{Benavoli2014} is also evaluated. The results from this Bayesian test are not included as they are practically indistinguishable from those for the regular signed-rank test. The $t$ and signed-rank tests are performed using \texttt{t.test} and \texttt{wilcox.test} from the \texttt{stats} package in \texttt{R}. The sign test is a special case of the binomial test, performed using \texttt{binom.test} from the same package. To handle data points equal to the null value $m_0$ (so-called ties) that can arise in testing discrete distributions, a modification to the sign test due to \textcite{Fong2003} is used. The \texttt{wilcox.test} function uses a normal approximation, which is capable of dealing with ties. The Bayesian bootstrap used in the familial test has the advantage of being insensitive to ties. The number of Bayesian bootstraps is fixed at $B=1,000$.

\begin{figure}[ht!]
\centering
\includegraphics{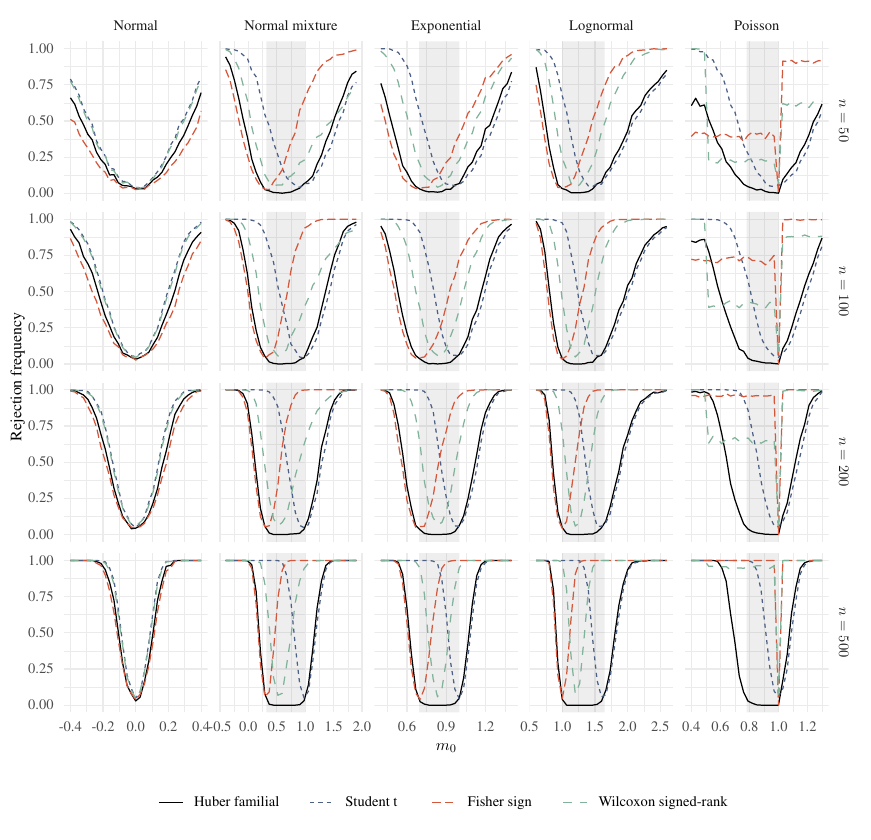}
\caption{Rejection frequency as a function of the null value $m_0$ in the one-sample (paired samples) setting. The sample size $n=200$. The shaded region indicates values of $m_0$ consistent with the familial null. Rejection frequency inside this region is size according to the familial null, and rejection frequency outside this region is power according to the familial alternative.}
\label{fig:one-sample}
\end{figure}

Figure~\ref{fig:one-sample} reports rejection frequencies for different values of $m_0$ as averaged over 1,000 simulations. The sample size is varied between $n=50$ and $n=500$. The shaded region indicates values of $m_0$ for which the familial (Huber) null is true. Rejection frequency inside this region indicates the size of a test according to the familial null. Power of a test according to the familial alternative is the rejection frequency outside this region. The frequentist tests are carried out at the 0.05 level. The familial test is conducted using loss matrix~\eqref{eq:lossmat}, which rejects when the null has posterior probability less than 0.05.

For the normal distribution, the familial test behaves similarly to the other tests. It has size no greater than 0.05 at $m_0=0$ and rejects sufficiently large departures from zero with high probability. Its power curve sits between those of the sign test and the signed-rank and $t$ tests. The $t$ test is well known to have optimal power here.

The story is more interesting for the normal mixture, exponential, and lognormal distributions. Here, the curves for the sign and $t$ tests attain their minima at different values of $m_0$ since the null of each test is true at different locations. The signed-rank test fails as a test of the median due to $X$ being asymmetric. The familial test behaves more conservatively than all three of these tests. It respects the familial null by rejecting with probability at most 0.05 in regions where some Huber centre is equal to $m_0$. In regions with no Huber centre equal to $m_0$, the familial test can be more powerful than the $t$ or sign tests. For instance, it is more powerful than the $t$ test for the exponential distribution when $m_0>1$. It is also more powerful than the sign test when $m_0<0.7$.

The Poisson distribution also tells an intriguing story. Since the Poisson is discrete, the power curves of the sign and signed-rank tests are step functions. In contrast to the other distributions, the curve of the sign test does not straddle the lower boundary of the familial null due to the lower boundary being some centre other than the median. The familial test respects its null and has good power for $m_0>1$ compared with the $t$ test.

\subsection{Independent samples}

\begin{figure}[ht!]
\centering
\includegraphics{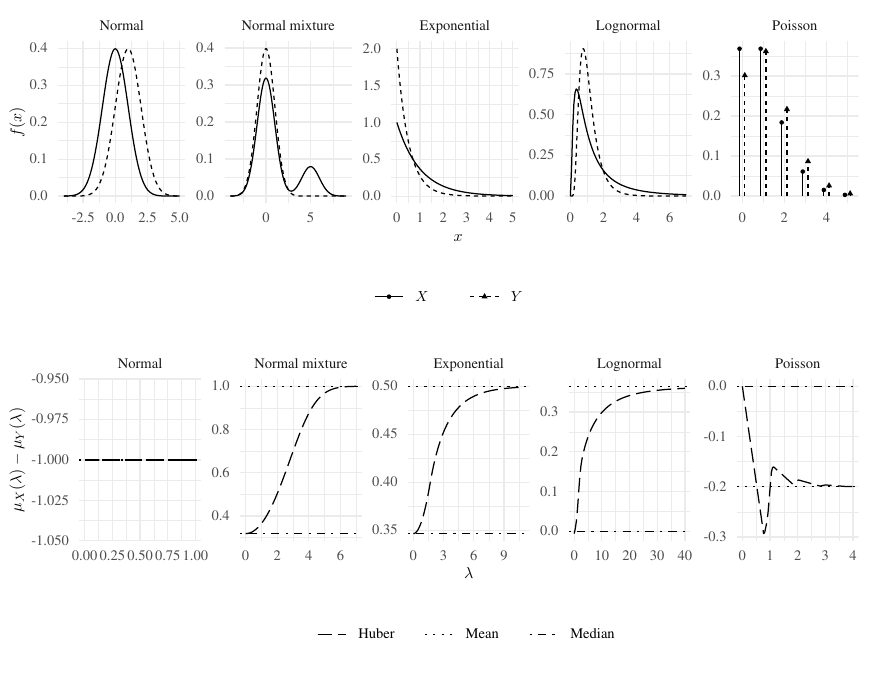}
\caption{Distributions analysed in the independent samples setting. The plots in the top row depict the density or mass function for the populations. The plots in the bottom row depict the corresponding difference in Huber families.}
\label{fig:two-sample-families}
\end{figure}

We now consider the independent samples setting with $X_1,\ldots,X_{n_1}$ and $Y_1,\ldots,Y_{n_2}$. The distributions analysed are: $X\sim\mathrm{Normal}(0,1)$, $Y\sim\mathrm{Normal}(1,1)$; $X\sim0.8\cdot\mathrm{Normal}(0,1)+0.2\cdot\mathrm{Normal}(5,1)$, $Y\sim\mathrm{Normal}(0,1)$; $X\sim\mathrm{Exponential}(1)$, $Y\sim\mathrm{Exponential}(2)$; $X\sim\mathrm{Lognormal}(0,1)$, $Y\sim\mathrm{Lognormal}(0,0.5)$; and $X\sim\mathrm{Poisson}(1)$, $Y\sim\mathrm{Poisson}(1.2)$. The distributions for $X$ are the same as those in the one-sample setting. Figure~\ref{fig:two-sample-families} plots the distributions and corresponding differences in Huber families.
For the normal, $Y$ is a location shift on $X$, so the difference in families is a singleton. For the remaining distributions, $Y$ has different skew and tailedness than $X$, so the difference in families are intervals. The Poisson is an example where the lower endpoint of the interval is not equal to the difference of means or medians.

We evaluate independent sample versions of the tests studied previously, summarized in Table~\ref{tab:two-sample-tests}.
\begin{table}[ht!]
\centering
\small
\begin{threeparttable}
\begin{tabular}{llll}
\toprule
Test & Null hypothesis ($\mathrm{H}_0$) & Alternative hypothesis ($\mathrm{H}_1$) & Center \\
\midrule
Huber familial & $\exists\,\lambda\in\Lambda:\mu_X(\lambda)-\mu_Y(\lambda)=m_0$ & $\forall\,\lambda\in\Lambda:\mu_X(\lambda)-\mu_Y(\lambda)\neq m_0$ & Huber \\
Welch $t$ & $\mu_X-\mu_Y=m_0$ & $\mu_X-\mu_Y\neq m_0$ & Mean \\
Mood median & $\mu_X-\mu_Y=m_0$ & $\mu_X-\mu_Y\neq m_0$ & Median \\
Wilcoxon rank-sum & $\mu_X-\mu_Y=m_0$ & $\mu_X-\mu_Y\neq m_0$ & Median\tnote{*} \\
\bottomrule
\end{tabular}
\begin{tablenotes}\footnotesize
\item [*] Provided $X$ and $Y$ only differ in location
\end{tablenotes}
\end{threeparttable}
\caption{Tests evaluated in the independent samples setting.}
\label{tab:two-sample-tests}
\end{table}
A Bayesian version of the rank-sum test by \textcite{Benavoli2015} is also evaluated. The results from that test are not materially different from those for the regular rank-sum test, so they are not reported. The $t$ and rank-sum tests are performed using \texttt{t.test} and \texttt{wilcox.test}. The median test is a special case of the chi-square test, performed using \texttt{chisq.test} from \texttt{stats}. Ties are again handled by \texttt{wilcox.test} via the normal approximation. For the median test, ties are discarded when calculating the test statistic.

\begin{figure}[ht!]
\centering
\includegraphics{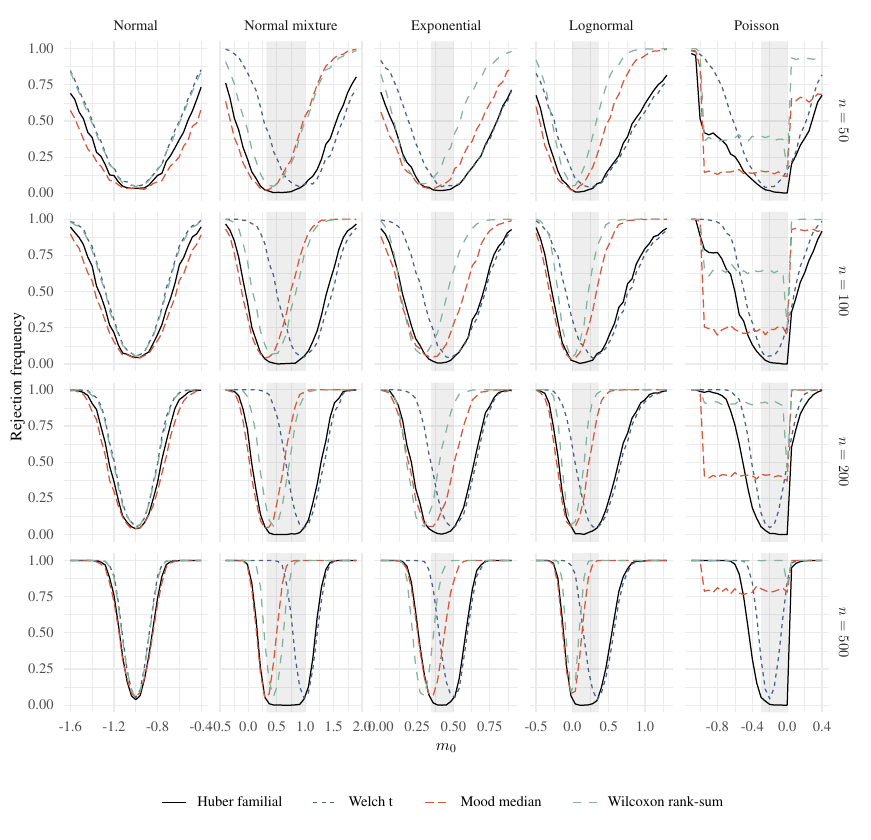}
\caption{Rejection frequency as a function of the null value $m_0$ in the independent samples setting. The sample sizes $n=200$. The shaded region indicates values of $m_0$ consistent with the familial null. Rejection frequency inside this region is size according to the familial null, and rejection frequency outside this region is power according to the familial alternative.}
\label{fig:two-sample}
\end{figure}

Results from 1,000 simulations are reported in Figure~\ref{fig:two-sample}. The shaded region again represents values of $m_0$ consistent with the familial null.

The power curves for the normal distribution are not too different from the one-sample setting. The Huber centre for $Y$ in the population is a point, so an independent samples test is not substantially different from a one-sample test with a point null.

For the normal mixture and exponential distributions, the rank-sum test fails as a test of medians since $X$ and $Y$ differ in scale, though curiously, it does not fail as a test of medians for the lognormal distribution. The $t$ and median tests reject at rates above 0.05 in the middle of the familial null region, where the difference in means and difference in medians are both far from $m_0$. There remains another Huber centre, not equal to the mean or median, for which the difference in centres is equal to $m_0$. The familial test accounts for this centre and correctly accepts the null with high probability.

The median test applied to the Poisson distribution does not have the correct size at $m_0=0$ due to ties in the data. Likewise, the rank-sum test fails as a test of medians for the Poisson due to $X$ and $Y$ differing in shape and scale. The power curve of the $t$ test does not straddle a boundary of the familial null. Unlike the median and rank-sum tests, the familial test succeeds as a test of medians, having zero size at $m_0=0$.

\end{appendices}

\end{document}